\documentclass[5p,twocolumn,10pt,times]{elsarticle}
\usepackage{amsmath}
\usepackage{hyperref}
\usepackage[ruled]{algorithm2e} 
\SetAlFnt{\scriptsize\sf}

\def\E{\mathbb{E}}
\def\R{\mathbb{R}}
\def\Z{\mathbb{Z}}
\def\N{\mathbb{N}}

\def\p#1{{\bf #1}}
\def\T#1{{\bf #1}}
\def\vet#1{{\left(\begin{array}{cccccccccccccccccccc}#1\end{array}\right)}}
\def\mat#1{{\left(\begin{array}{cccccccccccccccccccccccccccc}#1\end{array}\right)}}

\newtheorem{example}{Example}

\begin{document}
\baselineskip11pt

\begin{frontmatter}

\title{Topological computing of arrangements with (co)chains}

\author{Alberto Paoluzzi, Vadim Shapiro, Antonio DiCarlo, Francesco Furiani, Giulio Martella, Giorgio Scorzelli}
\cortext[mycorresponingauthor]{Corresponding author}
\address{Roma Tre University, University of Wisconsin-Madison \& ICSI, CECAM-IT-SIMUL Node, Scientific Computing and Imaging Institute (SCI)}

\begin{abstract}
In many areas of applied geometric/numeric computational mathematics, including geo-mapping, computer vision, computer graphics, finite element analysis, medical imaging, geometric design, and solid modeling, one has to compute incidences, adjacencies and ordering of cells, generally using disparate and often incompatible data structures and algorithms. This paper introduces computational topology algorithms to discover the 2D/3D space partition induced by a collection of geometric objects of dimension 1D/2D, respectively. Methods and language are those of basic geometric and algebraic topology. Only sparse vectors and matrices are used to compute both  spaces and maps, \emph{i.e.}, the chain complex, from dimension zero to three.
\end{abstract}


\end{frontmatter}


\section{Introduction}\label{introduction}

Given a collection \(\mathcal{S}\) of geometric
objects\footnote{Examples include, but are not limited to: line segments, quads, triangles, polygons, meshes, pixels, voxels, volume images, B-reps, \emph{etc.} 
In mathematical terms, a geometric object is a topological space embedded in some $\E^d$ \cite{Edelsbrunner:95}.},
the subject of this paper is computing the topology of their space
arrangement \(\mathcal{A}(\mathcal{S})\) as a \emph{chain complex},
\emph{i.e.},~as a short exact sequence of linear spaces \(C_i\) of (co)chains and linear
boundary/coboundary maps \(\partial_p\) and
\(\delta_p=\partial_{p+1}^\top\) between them: 
\[ 
C_\bullet = (C_p, \partial_p) := 
C_3 \ 
\substack{
\delta_2 \\
\longleftarrow \\[-1mm]
\longrightarrow \\
\partial_3 
}
\ C_2 \ 
\substack{
\delta_1 \\
\longleftarrow \\[-1mm]
\longrightarrow \\
\partial_2 
}
\ C_1 \ 
\substack{
\delta_0 \\
\longleftarrow \\[-1mm]
\longrightarrow \\
\partial_1 
}
\ C_0 .
\] 

The $C_\bullet$ chain complex fully characterizes the topology of the
cellular arrangement induced within the Euclidean space by a collection
of geometric objects embedded in it. The data structures needed for such
computational program are sparse 
multi-arrays, and their standard algebraic operations. 
In this way, we introduce a novel approach based on piecewise-linear algebraic topology~\cite{hatcher:2002,Rourke:Sanderson:1982}, that allows us to treat rather  general cellular complexes, with cells homeomorphic to polyhedra, \emph{i.e.},~to triangulable spaces~\cite{Rourke:Sanderson:1982}, and hence possibly non convex and multiply connected. 

We believe that basic geometric algebraic topology 
provides the right set of mathematical concepts and tools to compute
and explore the cells of the space partition induced by a set of
geometric objects, as well as the related incidence/neighborhood relations. 
The current escalation of quantity/quality of data, and their drift 
through pipelines of micro/macro-services that need simple interfaces, along with the fast diffusion of hybrid architectures for advanced applications, contribute to
motivate this paper. 
The notions we deal with include
geometric complexes, linear spaces of chains and cochains, the \emph{chain
complex} of linear maps between pairs of spaces, and their
compositions. The discussion is restricted to piecewise-linear topology
and to space dimensions less or equal to three. The paper formalizes the
algorithms to generate the matrices of (co)boundary operators on a
cellular arrangement in $\E^d$ space, with $d\in\{2,3\}$.

To our knowledge, algorithms for chain complex computing  only include  the recent paper~\cite{Alayrangues2015}, where combinatorial generalized maps~\cite{Lienhardt:2014}, a quite intricate data structure, are used for cycles/boundary calculation of the homology of a cellular complex.  Is is conversely well known~\cite{Munkres:84,Kannan:79,Paoluzzi:1993:DMS:169728.169719,Edelsbrunner:95} that for simplicial  complexes, \emph{i.e.},~triangulations, boundary operators are defined as linear extensions of basic boundary operators which act on simplices. 

Construction of arrangements of lines, segments, planes and other geometrical objects is discussed in~\cite{fhktww-a-07}, with a description of CGAL software~\cite{Fabri:2000:DCC:358668.358687}, implementing 2D/3D arrangements with Nef polyhedra~\cite{Hachenberger:2007:BOS:1247750.1248141,bieri:95}. A review of papers and algorithms concerning construction and counting of cells may be found in the chapter on Arrangements in the `Handbook of Discrete and Computational Geometry'~\cite{Goodman:2017:HDC:285869}. 
Arrangements of polytopes, hyperplanes and $d$-circles are discussed in~\cite{Ziegler:92}. 

The standard way to look at combinatorial data structures is the IG 
(Incidence Graph) data structure  described in the book
`Algorithms in Combinatorial Geometry' \cite{Edelsbrunner:1987:ACG:28905}.
IG is an implementation of the Hasse diagram~\cite{Birkhoff:1948} of the cells of a
\(d\)-complex \cite{ieee-tase}. Our (co)chain complex provides an
algebraic representation of the Hasse diagram with sparse matrices,
associating each two adjacent levels with a (co)boundary
map to traverse up and down the hierarchy.

Some early papers are concerned with the efficient representation of 3D
cellular decompositions~\cite{Baumgart:1972:WEP:891970,Muller:78}. In particular,
\cite{Dobkin:1987:PMT:41958.41967} defines the polygon-edge data
structure to represent orientable and contractible 3D decompositions
and their duals. Several other systems have been developed about three
decades ago, to handle the merging of complexes in the context of solid
modeling and manufacturing automation.  The merging of intersecting shells
of polyhedral solids is computable by finding all intersecting boxes of faces in $O(n\, log^2(n)+ k)$, where $n$ is the number of boxes and $k$ is the number of box-pair intersections~\cite{Hoffmann:1987:RSO:866286}.
The Selective Geometric Complex (SGC)
was introduced~\cite{Rossignac:89}  in 1989, allowing for lower dimensional cells contained inside the interior of cells. Later, scientists in UK proposed the Djin API
\cite{Woodwark:99,Middleditch:92}, with the theoretical
foundations for merging operation of cellular complexes, but (non-manifold) boundary representations remained the standard in the field.

Recently, a systematic procedure has been proposed in~\cite{Zhou:2016:MAS:2897824.2925901} for constructing a family of exact constructive solid geometry
operations, starting from a collection of boundary triangular meshes.

Most of earlier algorithms and procedures~\cite{4055948,
Ala:1992:PAB:616022.617736,
Baumgart:1972:WEP:891970,
bowyer1995introducing,
bowyer1995svlis,
Braid:1975:SSB:360715.360727,
Brisson:1989:RGS:73833.73858,
cadanda,
Dobkin:1987:PMT:41958.41967,
Gomes:1999:MMB:304012.304039,
Guibas:1985:PMG:282918.282923,
HoffmannK01,
ieee-tase,
Kalay:1989:HET:63718.63719,
Lee:2001:PES:376957.376976,
Lienhardt:1991:TMB:115604.115610,
Mantyla:1988:ISM:60949,
Paoluzzi:1989:BAO:70248.70249,
Paoluzzi:1993:DMS:169728.169719,
Paoluzzi:1995:GPP:212332.212349,
Pascucci:1995:DCB:218013.218055,
Pratt94ashape,
Raghothama:1999:CUD:304012.304019,
Rap97,
Requicha:1980:RRS:356827.356833,
RequichaVoelcker:77,
Rossignac:1991:CNG:115604.115606,
Rossignac:SGC:90,
Shapiro:1991:RSS:124951,
Shapiro:1995:PFS:218013.218029,
Silva:81,
Weiler:86,
Weiler:88,
Woo:85,
wozny1990geometric,
Yamaguchi:85,
yamaguchi1995ntb,
Zhou:2016:MAS:2897824.2925901,
bieri:95,
Rossignac:89,
Hoffmann:91,
Hoffmann:1989:GSM:74803,
Hoffmann:1987:RSO:866286} 
work with data
structures optimized for selected classes of geometric
objects. Contrary-wise, our formulation, cast in
terms of (co)chain complexes of (co)boundary maps, may be
applied to very different geometric objects, ranging from solid models
to engineering meshes, geographical systems, biomedical images.

Numerical methods aiming to integrate domain modeling, differential
topology and mathematical modeling with physical simulations are also based
on chains and cochains~\cite{PALMER1995733,Palmer1993}. In particular, Discrete Exterior Calculus
(DEC) with simplicial complexes was introduced by
\cite{Hirani:2003:DEC:959640} and made popular by
\cite{Desbrun:2006:DDF:1185657.1185665,Elcott:2006:BYO:1185657.1185666}. 
FEEC is a recent advance~\cite{arnold_falk_winther_2006,Arnold:2010,Arnold:2018} in the mathematics of finite element methods that employs differential complexes to construct stable numerical schemes.
The Cell Method~(CM) 
is a purely algebraic computational method 
for modeling and simulation~\cite{Tonti:1975,Tonti:2013,Ferretti:2014} based on boundary/coboundary maps and
 a direct discrete formulation of field laws.
Our own research in geometrical and physical modeling  with chain and cochain complexes was introduced in \cite{DiCarlo:2009:DPU:1629255.1629273,ieee-tase,Dicarlo:2014:TNL:2543138.2543294}.

All merge algorithms, by nature, follow the same steps. The
advantages of formulating them in terms of (co)chain complexes and operations on sparse matrices are that (1)  the common and general algebraic topological
nature of this operation is revealed; (2) implementation specific low-level
details and algorithms are hidden; (3) explicit connection to SpMV kernels (sparse matrix-vector  multiplication) and to sparse
numerical linear algebra systems~\cite{Bell:SpMV:NVIDIA:2008,gemmexp,Davis:2006:DMS:1196434,BEKS14}  
on GPU and HPC platforms is provided; (4) systematic and rigorous development of the
algorithms, that are correct by construction, is supported.

\begin{figure}[h] 
   \centering
   \includegraphics[height=0.43\linewidth,width=0.5\linewidth]{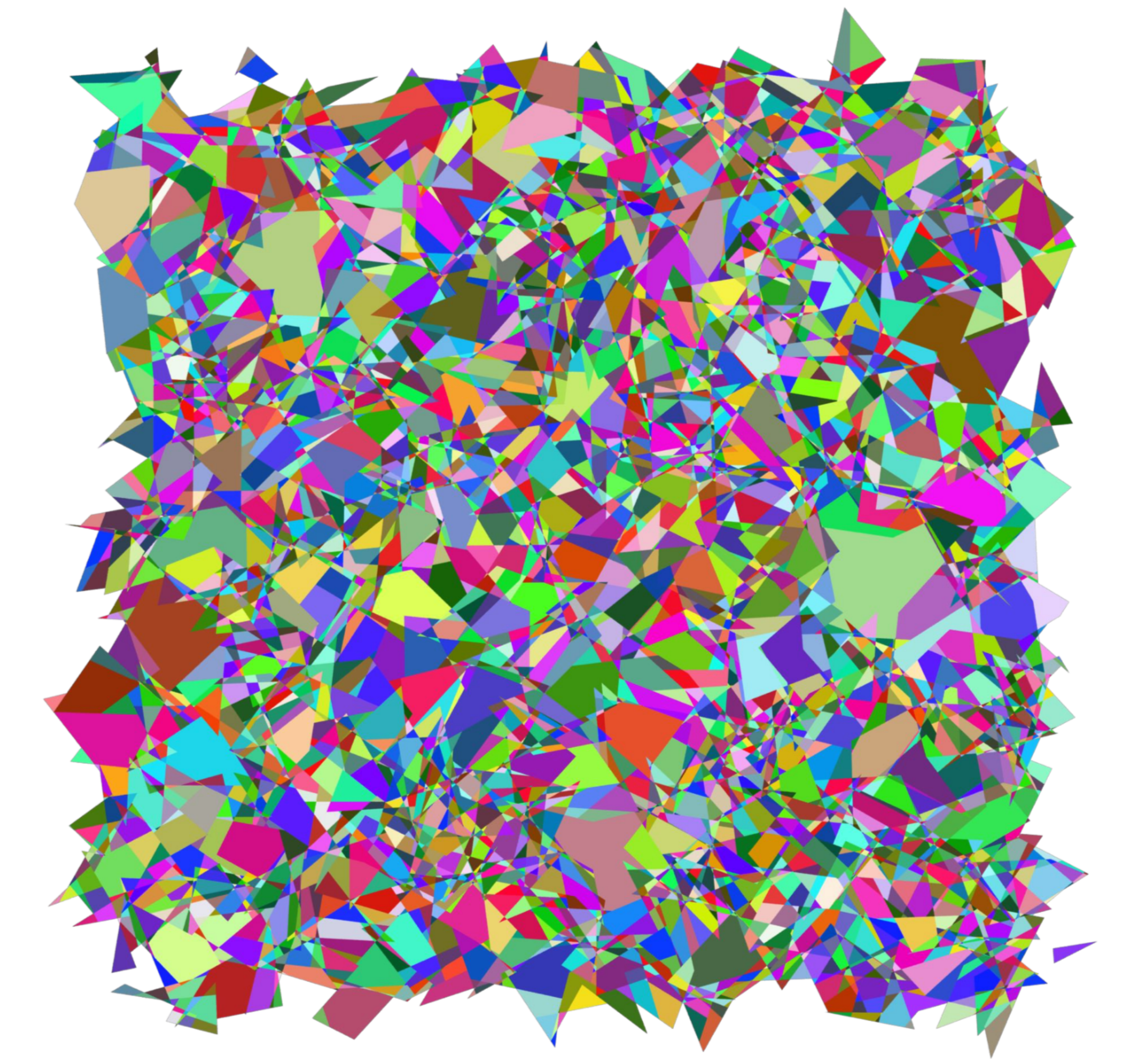}%
   \includegraphics[height=0.43\linewidth,width=0.5\linewidth]{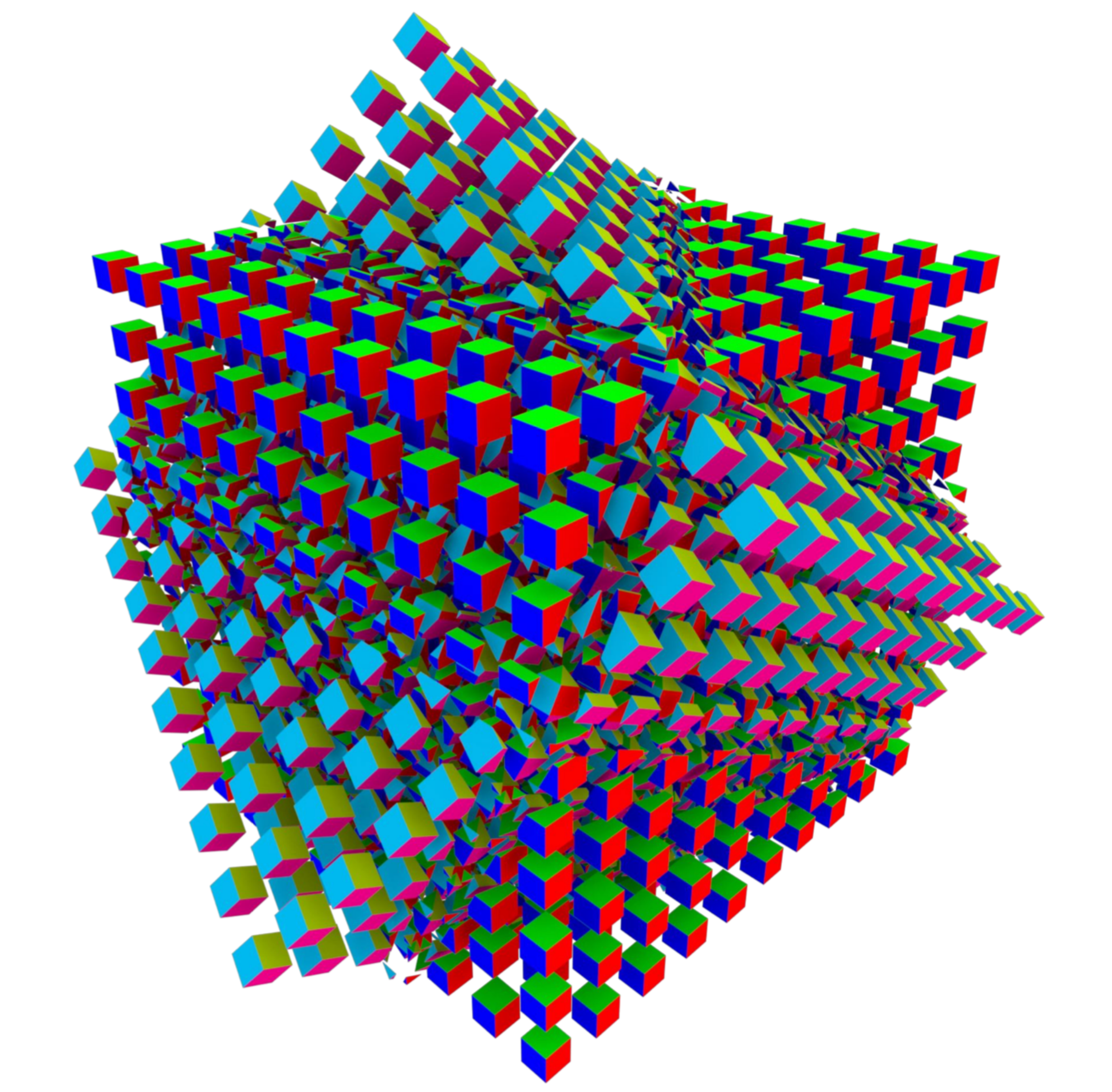}%
   
   {\footnotesize\hspace{.23\textwidth}(a)\hfill(b)\hspace{.23\textwidth}}
   \vspace{-2mm}
   \caption{ Examples:
   (a) Regularized 2D arrangement $X_2$ of the plane generated by a set of random line segments. The Euler characteristic is $\chi = \chi_0 - \chi_1 + \chi_2 = 11361 - 20813 + 9454 = 2$; (b) merge of two 3-complexes with $2\times 10^3$ 3-cells. The Euler characteristic (of the non-exploded resulting 3-complex) is $\chi = \chi_0 - \chi_1 + \chi_2 - \chi_3 = 8787 - 26732 + 26600 - 8655 = 0$. This  count includes the outer (unbounded) 2-cell or 3-cell, respectively, that are also computed by theTopological Gift Wrapping  algorithm TGW (see Section~\ref{tgw-algorithm}). 
   The Euler characteristic of the $d$-sphere is $\chi = 1 + (-1)^d =$ 2 or 0 for either even or odd space dimension $d$. }
   \label{fig:arxiv}
\end{figure}

The paper is organized as follows. A brief introduction to the proposed 
computational pipeline is given in Section \ref{sec:pipeline}, going
from the arrangement of \(\E^2\) induced by a set of line segments, to
the arrangement of \(\E^3\) induced by a collection of open/closed
piecewise-linear surfaces and/or meshes. 
In the short subsection~\ref{sec:prologue}, we show that representing chains as sparse arrays is compact and flexible.
In Section \ref{sec:algorithms}, 
the algorithms for computing (co)boundary operators are presented in pseudocode format. In Section \ref{other-works}, our approach is compared with the relevant literature, remarking differences and merits. Past development and current prospects of this project are outlined in Section \ref{sec:projects}. The closing section presents a summary of contents, and outlines 
 possible applications of ideas. 
The Appendix  gives the theoretical minimum on chains and examples of topology computation.

\section{Computational pipeline}
\label{sec:pipeline}

Let us start with an input collection $\mathcal{S}$ of piecewise-linear geometric objects of  $(d-1)$ dimension, embedded in $\E^d$ space, with $d \in \{2,3\}$.  Examples include soups of lines or  polygons, triangled surfaces,  quads from cubical meshes, 1-, 2-, or 3-cells from 2D or 3D image elements (pixels or voxels, respectively), 2-skeletons/boundaries of triangulated polyhedra, non manifold B-reps or decompositive reps of solid models. These objects are \emph{geometric complexes}, \emph{i.e.}, pairs $(X,\mu)$, where $X$ is a cellular complex\footnote{See Appendix~\ref{chain-complexes-1-page-definitions} for this and related definition(s).} specifying the topology and $\mu: X_0 \to \E^d$ is the embedding function of 0-cells, sufficient for a piecewise-linear geometry.
The data may contains both $(d-1)$- and $d$-complexes: the combinatorial union of their $(d-1)$-skeletons is selected as the actual input to the pipeline.
An admissible input collection $\mathcal{S}$ of geometric complexes will mutually intersect and partition $\E^d$ into a cellular complex $X = \bigcup X_p$ $(0\leq p\leq d)$, called the \emph{arrangement} $\mathcal{A}(\mathcal{S})$ induced by $\mathcal{S}$. 

The object of this paper is the computation of the chain complex $C_\bullet(X) = (C_p,\partial_p)$, starting from some representation\footnote{Our prototype implementation in \\ \href{https://github.com/cvdlab/LinearAlgebraicRepresentation.jl/tree/julia-1.0}{https://github.com/cvdlab/LinearAlgebraicRepresentation.jl/tree/julia-1.0}, makes use of LAR representation~\cite{Dicarlo:2014:TNL:2543138.2543294}, on which this approach strongly relies.} of $\mathcal{S}$. In particular, we compute the matrices of the linear maps $\partial_p$ (and their duals $\delta_{p-1}$)  between chain spaces $C_p$. Definitions and examples are given in Appendix~\ref{sec:appendix}.
Since the matrix of a linear map $C_p \to C_{p-1}$ between linear spaces contains by columns the target space representation of the domain space basis elements, the paper also provides constructive algorithms to generate a sparse matrix representation of basis elements $u_p\in C_p$, which are one-to-one with $p$-cells $\sigma_p$ in $X_p$ skeletons. Note that cells in $X$, specifically in $X_d$, are not known in advance. 

\begin{figure}[htbp] 
   \centering
   \includegraphics[width=\linewidth]{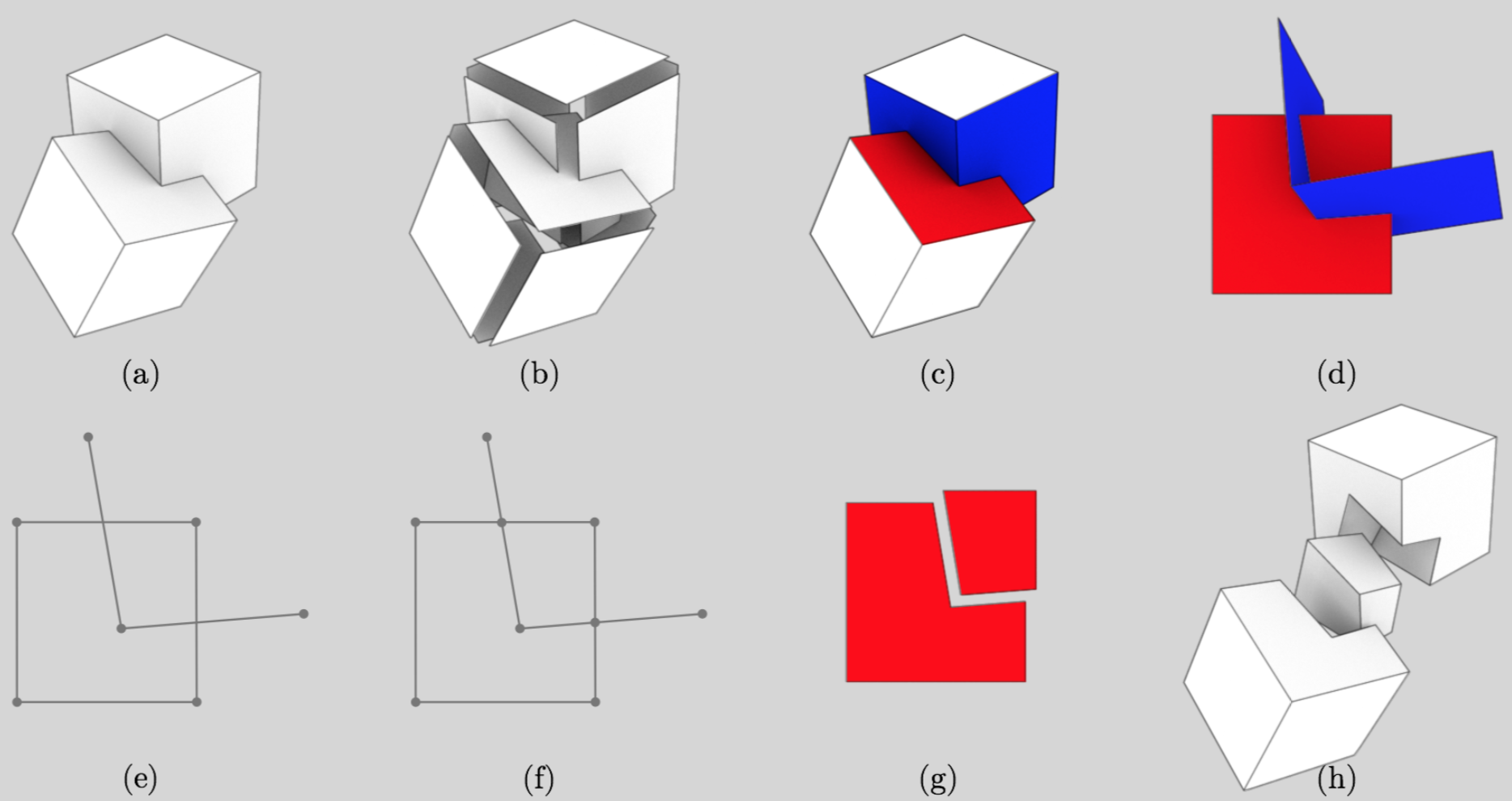}%
   
   \caption{Cartoon display of the computational pipeline: (a) two solids in $\mathcal{S}$; (b) the exploded input collection $\mathcal{S}_2$ in $\E^3$; (c) 2-cell $\sigma$ (red) and the set $\Sigma(\sigma)$ (blue) of possible intersection; (d)  $\sigma\cup\Sigma$ affinely mapped on $z=0$; 
   (e)~reduction to a set of 1D segments in $\E^2$ via intersection with $z=0$; (f) pairwise intersections; (g) exploded $U_2$ basis of $C_2$ generated as columns of $\partial_2: C_2 \to C_1$; (h) exploded $U_3$ basis of $C_3$ generated as columns of operator's $\partial_3: C_3 \to C_2$ sparse matrix, via the TGW algorithm in 3D.    }
   \label{fig:process}
\end{figure}

The computation is correct and robust  because the boundaries of adjacent decomposed 2-cells are compatible as cellular complexes by construction. This fact is induced by continuity of topological spaces, mathematically modeled here by chains of cells of a complex. A requirement of the standard definition of a cellular complex~\cite{Munkres:84,hatcher:2002} demands boundary compatibility to hold. This fact is guaranteed here, since abutting subsets of 1-cells have non-empty intersection, so they generate congruent 0-, 1-cells. 
A summary of the computational steps for $d=3$ follows.

\subsection{2D arrangements generated by 2-cells (Merge)}\label{facet-arrangements}

Let $\mathcal{S}_2 \subseteq \mathcal{S}$ be the set of 2-cells of input geometric complexes\footnote{Cell complex embedded in Euclidean space via association of position vectors to 0-cells.}, embedded in $\E^3$. 
Note that $\mathcal{S}_2$ is not required to be a cellular complex, since cells may intersect outside of their boundaries. 
It is only required that each cell is connected and manifold. Each $\sigma\in S_2$ is mapped to subspace $z=0$ by an affine transformation $\T{Q}_\sigma$, together with the set $\mathcal{I}(\sigma)\subset S_2$ of cells potentially intersecting it. The set $\Sigma = X_1(\sigma \cup \mathcal{I}(\sigma))$ is intersected with $x=0$, producing a set $\mathcal{S}_1(\sigma)$ of line segments in $\E^2$. 

\begin{figure}[htbp] 
   \centering
   \includegraphics[width=0.65\linewidth]{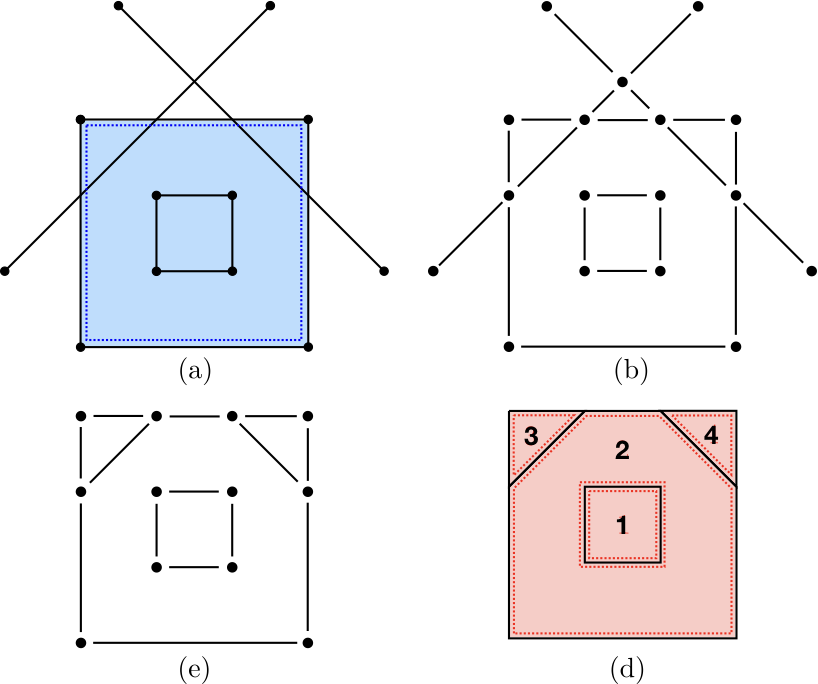} 
   \caption{Computation of the regularized arrangement of a set of lines in $\E^2$: (a) the input, \emph{i.e.},~the 2-cell $\sigma$ (signed blue) and the line segment intersections of $\Sigma(\sigma)$ with $z=0$; (b) all pairwise intersections of 1-cells; (c)~removal of the 1-subcomplex external to $\partial\sigma$; (d) the 2-chain generated by $\sigma\cup\Sigma$ via TGW in 2D.}
   \label{fig:2D}
\end{figure}

First, these are mutually intersected, producing the chain complex $C_\bullet(\sigma) = (C_{2,1,0},\,\partial_{2,1})$ generated by $\mathcal{A}(\mathcal{S}_1)$ (see Figure~\ref{fig:2D}d).
Care must be taken to identify the 1-cycles around holes within partitioned 2-cells, in order to remove their outer boundary cycles, by removing their columns, as well the outer cycle, from the operator matrix. Identification is easy: each hole produces two opposite columns summing to 0.
Finally, the geometric 2-complex $X_\sigma$ is transformed back in $\E^3$ by $\T{Q}^{-1}_\sigma$. 
In summary, Algorithm~\ref{alg:two} executes the one-to-one map $\sigma\mapsto X_\sigma$, by computing the maps $\sigma\mapsto C_\bullet(\sigma)$ independently from each other. The output is a  set $\varmathbb{C} := \{ C_\bullet(\sigma), \sigma\in \mathcal{S}_2\}$.

\subsection{Quotient set computations (Congruence)}\label{quotient-set}

The idea allowing us to compute independent fragmentations of 2-cells comes from a similitude between homology and congruence. Two $(d-1)$-spaces (curves, surfaces, \emph{etc.}) embedded in $\E^d$  are topologically \emph{homologous} when their boundaries can be glued, enclosing a portion of the ambient space, and subdivide $\E^d$ in two parts, inner and outer. 
Two geometric figures are geometrically \emph{congruent} iff one can be transformed into the other by an isometry~\cite{Coxeter:1967}. The congruences $R_p$ between $p$-cells of geometric complexes in $\varmathbb{C}$ are equivalence relations, so we may compute the chain complex of  quotient chain spaces, 
\[
C_2(U_2/R_2) \overset{\partial_2}{\longrightarrow} C_1(U_1/R_1) \overset{\partial_1}{\longrightarrow} C_0(U_0/R_0),
\]
over which subsequently build the yet unknown basis of $C_3$. Note that  $U_p = \bigcup\,U_p^\sigma$, with $\sigma\in\mathcal{S}_2$, is the union of bases of fragmented $p$-chains in $\varmathbb{C}$, module the congruence relations $R_p$. \ 
$C_p(U_p/R_p)$ stands for the chain space generated by $X_p = U_p/R_p$.
 In this stage of the computational pipeline, we compute, for each $\sigma\in\mathcal{S}_2$, the quotient sets and the maps $\partial_p$ in-between, for $p=0,1,2$. 

As usual, we proceed by  an inductive procedure: each stage consists in glueing cells of given dimension to the result of the previous stage~\cite{Baladze:EoM}. The construction of the sparse matrix of the signed operator $\partial_1: C_1\to C_0$ is straightforward: for each $u_1^h = u_0^{k_2} - u_0^{k_1}$, just write $\partial_1[k_2,h]=1$ and $\partial_1[k_1,h]=-1$, by convention for $k_2>k_1$. For details of quotient operations, see Section~\ref{quotients}.

\begin{example}[Merge with incompatible boundaries]
\label{ex:square}

Here we discuss the merge result when the input collection is defined by a pair of adjacent complexes with incompatible sub-complexes along their interface. Such a 2D example is displayed in Figure~\ref{fig:abutting}.
Another simple example may include two tetrahedralized unit cubes that are incident on a planar face triangulated into two triangles but along different diagonals. Similarly to the 2D case, each boundary triangle would be fragmented against all the incident ones, producing  fragmented output faces, so that the result on the common affine support (say, the vertical plane) would be exactly four triangles, because each input triangle is fragmented by the other diagonal. A larger  example is shown in Figure~\ref{fig:arxiv}a, where the 2D arrangement generated by a number of random line segments is displayed.
\begin{figure}[htbp] 
   \centering
   \includegraphics[width=\linewidth]{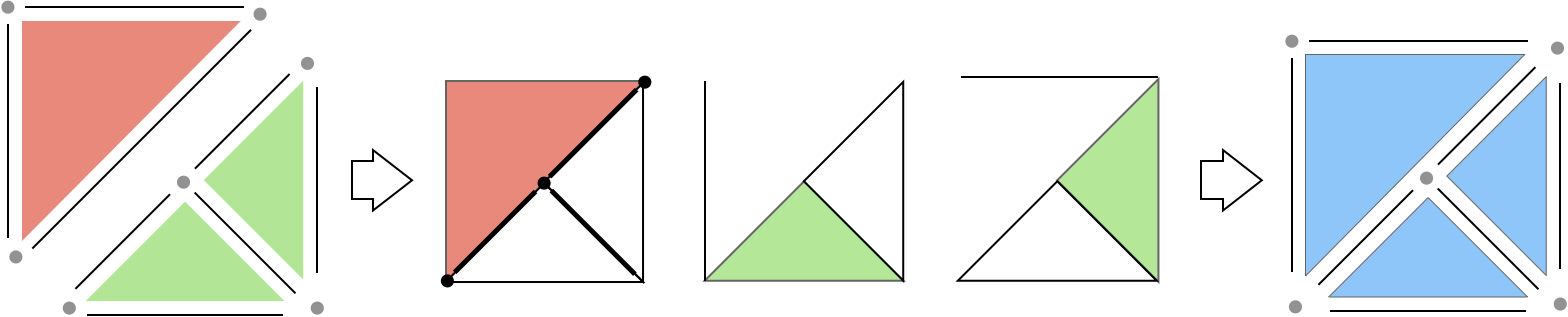}
   \\
   {\footnotesize \hspace{0.06\linewidth}(a)\hspace{0.38\linewidth}(b)\hspace{0.38\linewidth}(c)\hspace{0.04\linewidth}}\vspace{-3mm}
   \caption{Merge of two 2D complexes with incompatible boundaries: (a) Input data $\mathcal{S}_2 = \Sigma^1 \cup \Sigma^2$; (b)~independent fragmentation of 2-cells $\sigma\in\mathcal{S}_2$ induced by $\mathcal{I}(\sigma)$; (c)  \emph{local} arrangement $X_2 = \mathcal{A}(\mathcal{S}_1(\mathcal{S}_2))$ generated by \emph{Merge},  2D \emph{TGW}, and \emph{Congruence} algorithm pipeline. 
   }
   \label{fig:abutting}
\end{figure}
\end{example}
\vspace{-3mm}
Note in Figure~\ref{fig:abutting} that: (i) each 2-cell $\sigma$ is processed \emph{independently} as $\Sigma = \{\sigma\}\cup\mathcal{I}(\sigma)$; (ii) the bigger diagonal is fragmented by the normal 1-cell; (iii) the unit 2-chains are reconstructed in 2D (before reduction by congruence) as 1-cycles by the TGW algorithm; (iv) 0-cells and 1-cells from fragmented diagonal are finally identified \emph{mod congruence}.

\subsection{Topological gift wrapping (TGW)}\label{tgw-algorithm}

The  algorithm discussed here is used to compute topologically in 2D/3D, respectively, the sparse matrices of signed operators $\partial_2$ and $\partial_3$, starting from $\partial_1$ and $\partial_2$ input, respectively.  The matrix $[\partial_d]$ of the linear map $C_d \to C_{d-1}$ between linear spaces contains by columns the target space representation of domain space basis elements, as closed $(d-1)$-chains, \emph{i.e.},~$(d-1)$-cycles.
Within the computational pipeline discussed in this paper, TGW is used \emph{locally} for each 2-cell to be decomposed, and \emph{globally} to generate the 3-cells of the arrangement of the ambient  space $\E^3$. The algorithm pseudocode is given and discussed in Section~\ref{partial_3-computation}. The needed extensions for handling holes and non-connected components are detailed in Section~\ref{delta_2-computation}.

\begin{figure*}[htbp] 
   \centering
\includegraphics[width=0.2\textwidth]{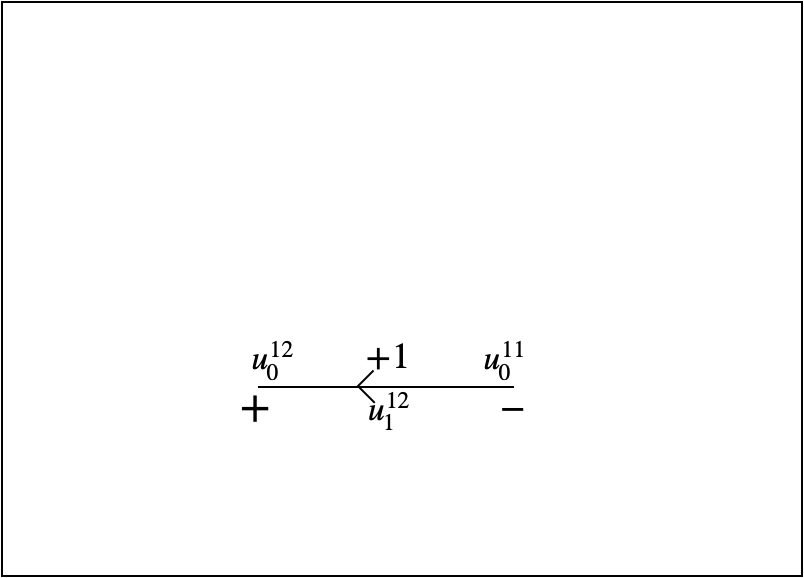}\includegraphics[width=0.2\textwidth]{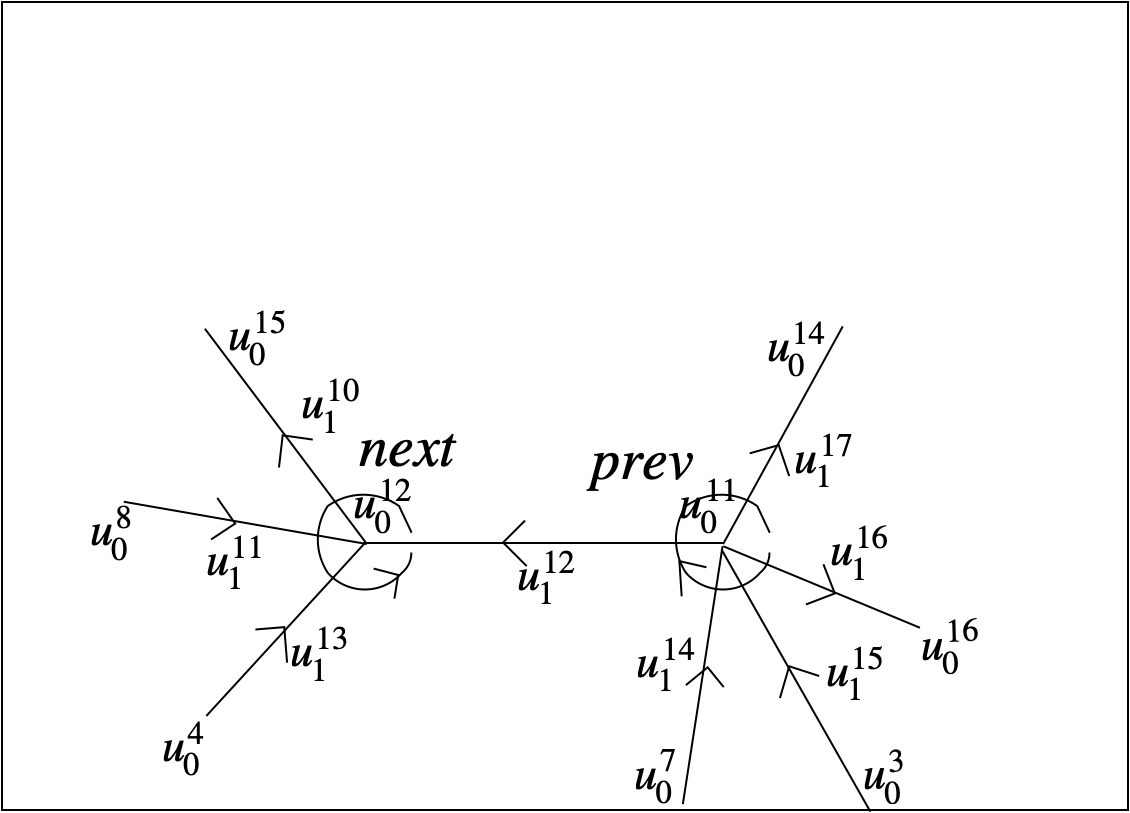}\includegraphics[width=0.2\textwidth]{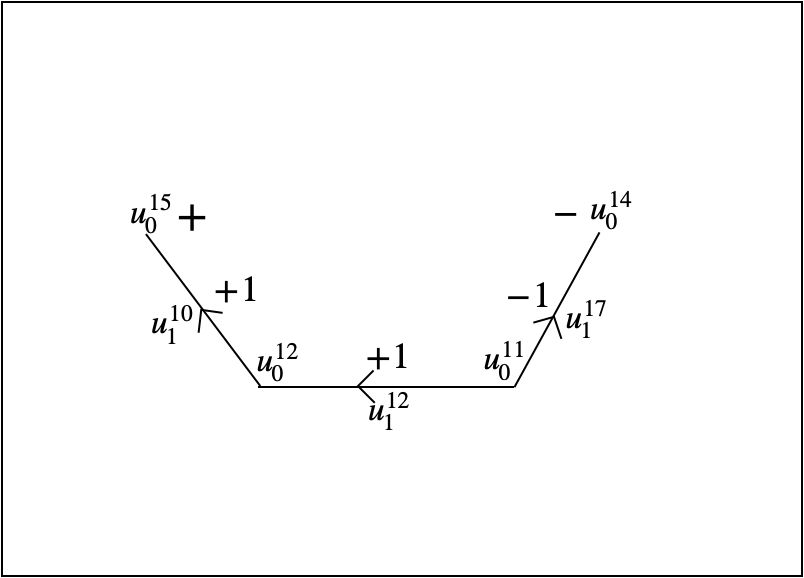}\includegraphics[width=0.2\textwidth]{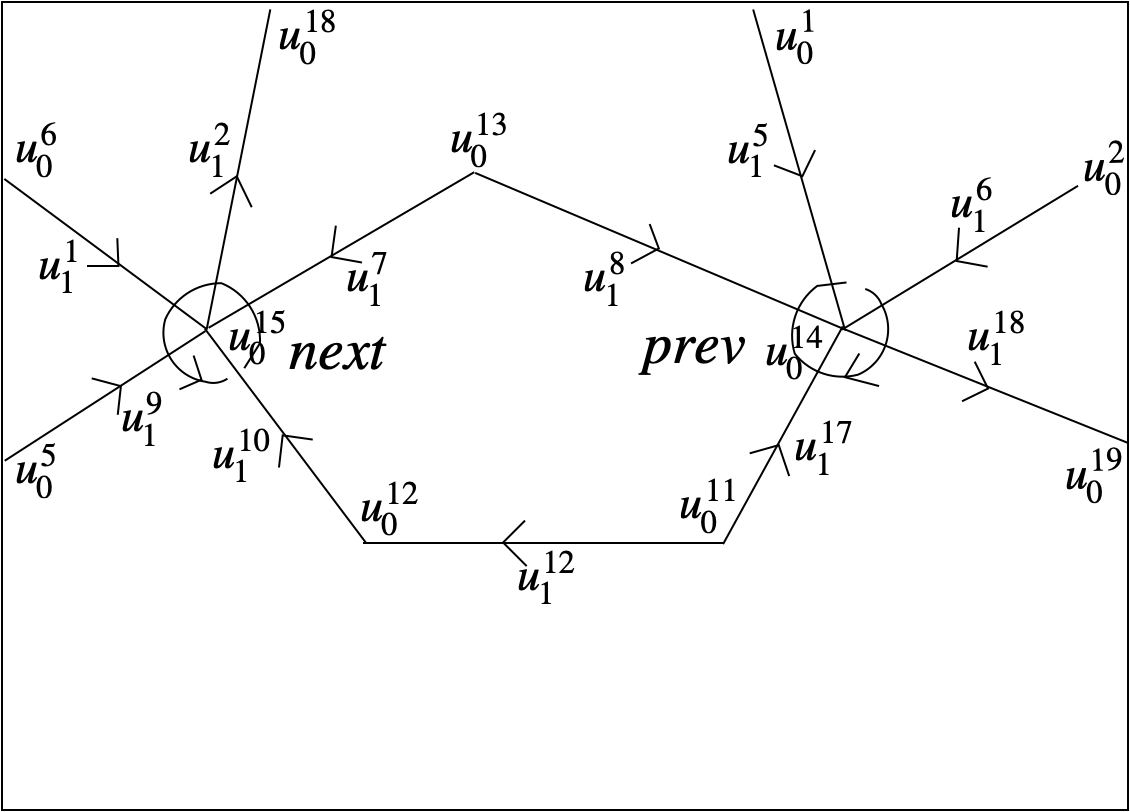}\includegraphics[width=0.2\textwidth]{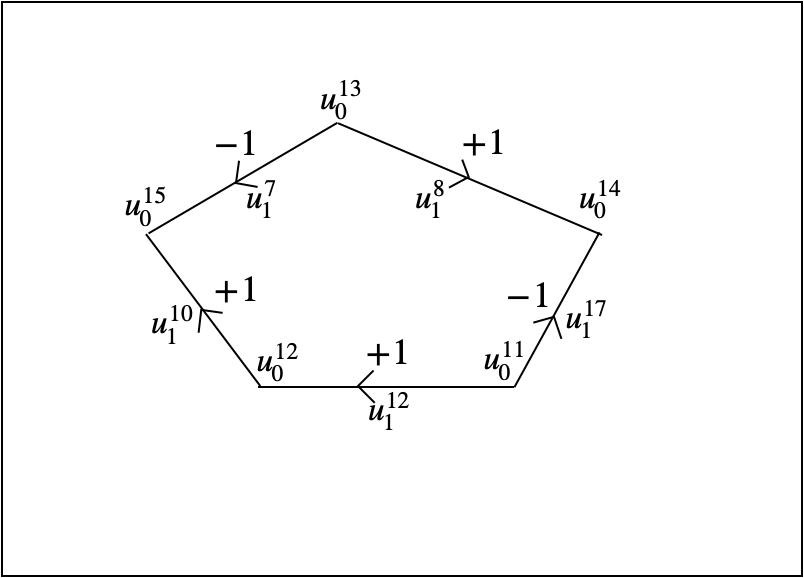}

{\footnotesize\hspace{.09\textwidth}(a)\hfill(b)\hfill(c)\hfill(d)\hfill(e)\hspace{.09\textwidth}}
\vspace{-2mm}   
\caption{Extraction of a minimal 1-cycle from $\mathcal{A}(X_1)$: (a) the initial value for $c\in C_1$ and the signs of its oriented boundary; (b) cyclic subgroups on $\delta\partial c$; (c) new (coherently oriented) value of $c$ and $\partial c$; (d) cyclic subgroups on $\delta\partial c$; (e) final value of $c$, with $\partial c = \emptyset$. The step-by-step computation is discussed in Example~\ref{example1}.}
\label{fig:step-by-step}
\end{figure*}

\begin{figure*}[htbp] 
\hfill\includegraphics[height=0.15\textwidth,width=0.1428\textwidth]{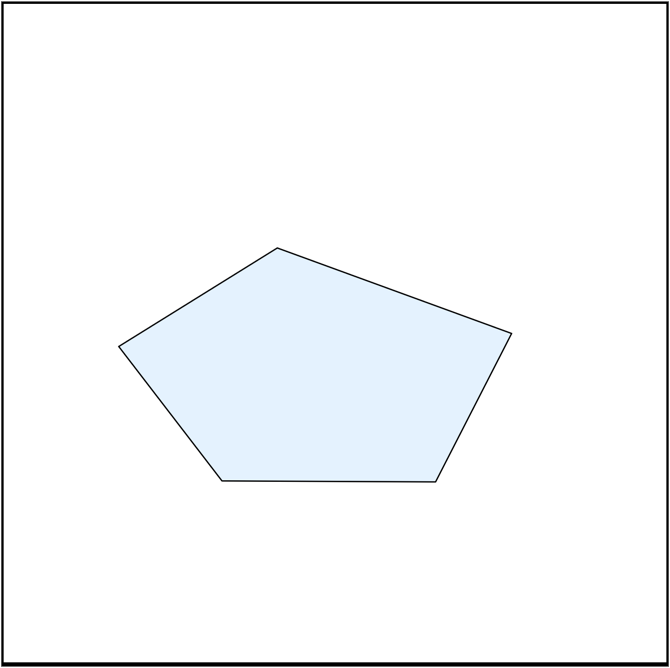}%
\includegraphics[height=0.15\textwidth,width=0.1428\textwidth]{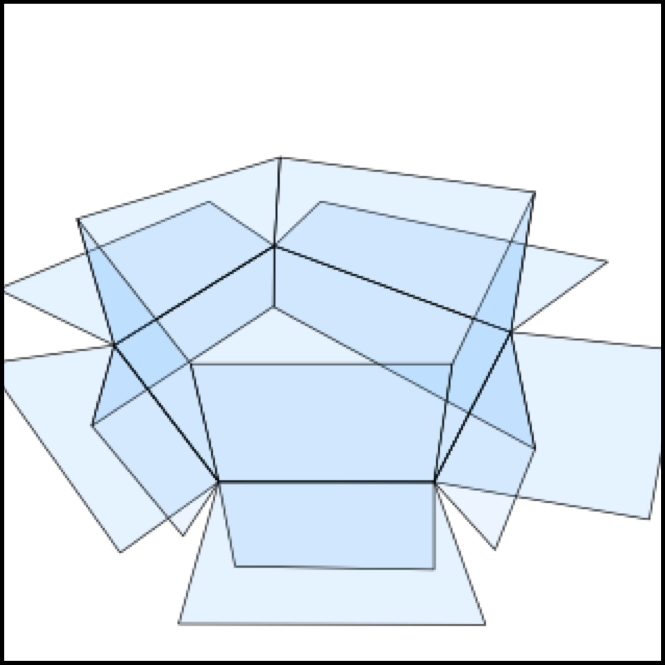}%
\includegraphics[height=0.15\textwidth,width=0.1428\textwidth]{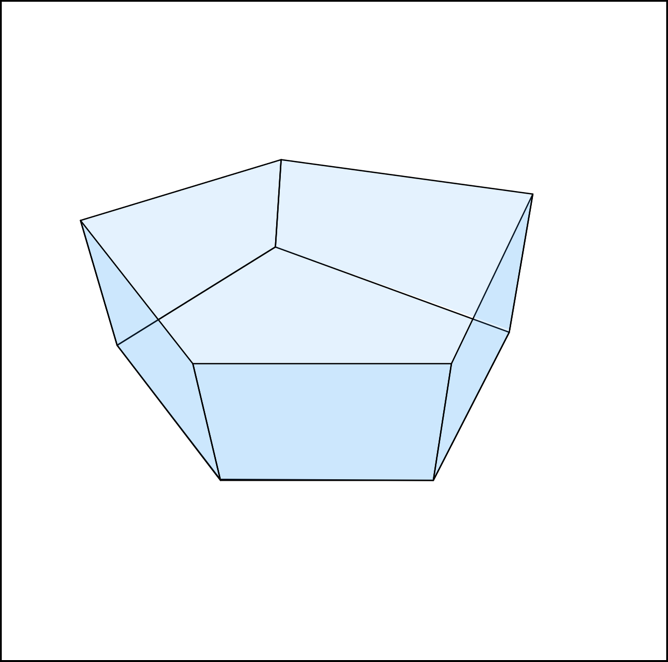}%
\includegraphics[height=0.15\textwidth,width=0.1428\textwidth]{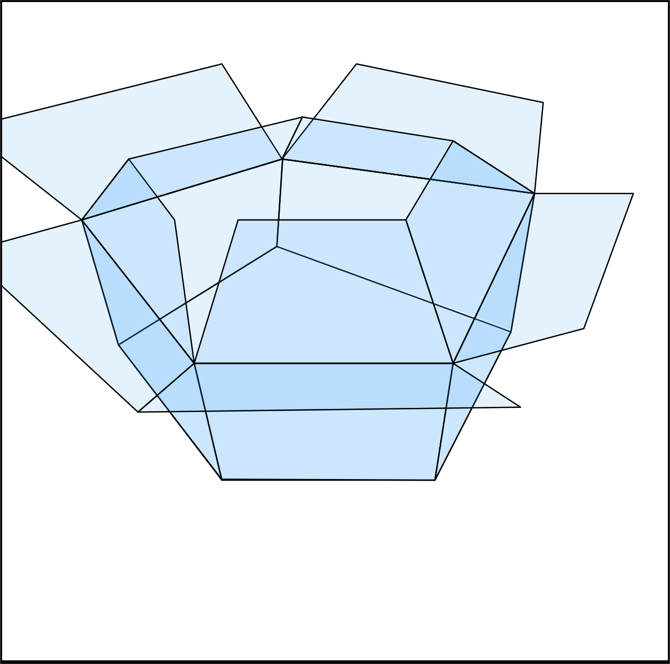}%
\includegraphics[height=0.15\textwidth,width=0.1428\textwidth]{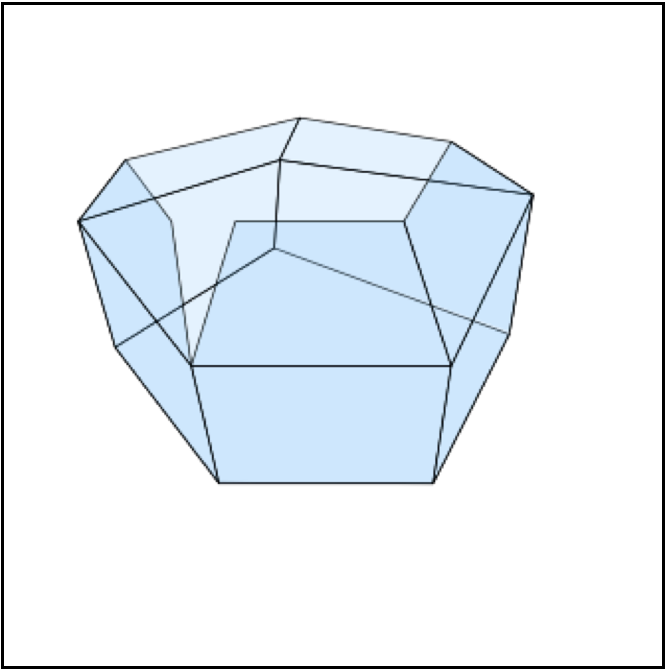}%
\includegraphics[height=0.15\textwidth,width=0.1428\textwidth]{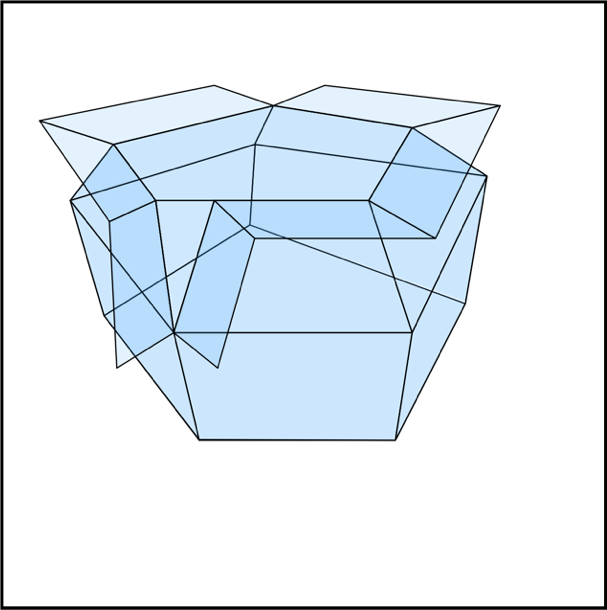}%
\includegraphics[height=0.15\textwidth,width=0.1428\textwidth]{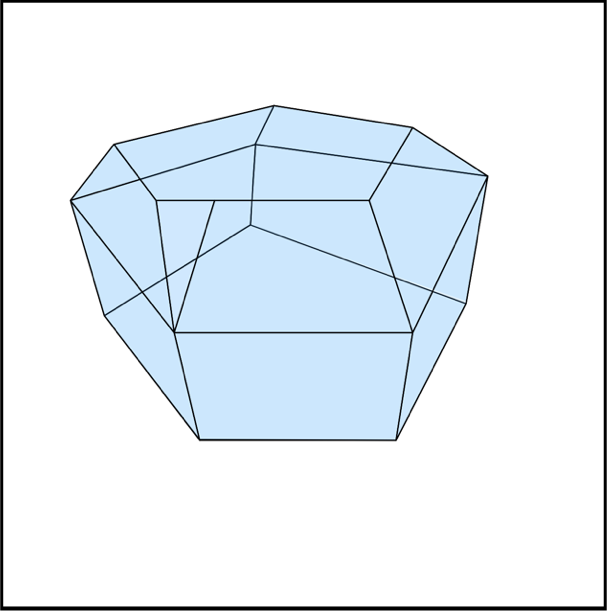}

{\footnotesize\hspace{.06\textwidth}(a)\hfill(b)\hfill(c)\hfill(d)\hfill(e)\hfill(f)\hfill(g)\hspace{.06\textwidth}}

\caption{Extraction of a minimal 2-cycle from $\mathcal{A}(X_2)$: (a) initial (0-th) value for $c\in C_2$; (b) cyclic subgroups on  $\delta\partial c$; (c)~1-st value of $c$; (d) cyclic subgroups on $\delta\partial c$; (e) 2-nd value of $c$; (f) cyclic subgroups on $\delta\partial c$; (g) 3-rd value of $c$, such that $\partial c=0$, hence stop.}
   \label{fig:3D}
\end{figure*}

The topological method introduced here, reminiscent of the ``gift-wrapping'' algorithm~\cite{Cormen:2009:IAT:1614191,Jarvis:1973:ICH} for computing convex hulls of 2D and 3D discrete sets of points, is detailed and formalized in Section~\ref{partial_3-computation}. The TGW algorithm takes a sparse matrix $[\partial_{d-1}]$ as input and produces in output a sparse matrix  $[\partial_{d}^+]$, augmented with the outer cell. A geometric embedding function $\mu: X_0\to\E^d$ is used to compute the angular ordering, around some $(d-2)$-cells, of $(d-1)$-basis elements in the boundary's coboundary, while wrapping up a $(d-1)$-cycle, as illustrated in Figures~\ref{fig:step-by-step} and \ref{fig:3D}. The built (minimal) cycles are set as columns of $[\partial_d]$, in the construction of the $C_d$ basis.
Note also that, once the ordered sets of basis elements is fixed, columns contain the \emph{coordinate representation} of $(d-1)$-cycles, built from group coefficients $(\{-1,0,1\},\,+) \simeq \Z/{3}\Z = \Z_3$. Analogously,  boundaries and coboundaries of chains are calculated by multiplication of operator matrices times proper coordinate vectors of such coefficients.

\subsection{Non-connected components (Holes)}

The outer cell of the space arrangement $X = \mathcal{A}(\mathcal{S})$ might have a non-connected boundary, compraising more than one $(d-1)$ cycle\footnote{Called a \emph{shell} in the literature of solid modeling.}, like a  3-ball minus a smaller concentric 3-ball. Analogously, $X$  might contain both non-connected and possibly nested components. The TGW algorithm actually computes all of the boundary cycles, that must be properly handled in order to produce a single $[\partial_d]$ matrix:  first decompose the input $[\partial_{d-1}]$ into connected components; then assemble/remove the empty cycles.

\subsubsection{Decomposition of 2-skeleton}\label{decomposition}

Consider a bipartite graph {$G = (N, A)$, with $N\simeq\Lambda_2 \cup \Lambda_0$, and $A \subseteq \Lambda_2 \times \Lambda_0$,} associated with the sparse characteristic matrix encoding the \emph{incidence} relation. $G$ has one node for each 2-cell, one node for each 0-cell, and one arc for each incident pair. Therefore, the arcs in $G$ are one-to-one with the nonzero elements of the $A$ matrix. By computing the {maximal} point-connected components of $G$, we subdivide the $X_2$ skeleton into $h$ connected components: $\mathcal{X}_2 = \{ {X}_2^p \}$, such that {$1\leq p\leq h$.}
For each component ${X}_2^p$, repeat the following actions. First, assemble the $[\partial_2]^p$ sparse matrix, and compute the corresponding $[\partial_3^+]^p$ generated by Algorithm~\ref{alg:one}. Then, subdivide $[\partial_3^+]^p$ into the boundary operator $\partial_3^p: C_3^p\to C_2$ and  the column matrix $c^p = \partial_3^+[\sigma^p] \in C_2$ of the outer cell $\sigma^p\in X_3$.
The set $S = \{ c^p \}$ of $h$ disjoint 2-cycles, is the initialization of the set of $X_d$ \emph{shells}. 
Other (empty) shells of $X_d$ can be discovered later from mutual containment of $S$ elements.

\begin{figure}[htbp] 
   \centering
   \includegraphics[width=\linewidth]{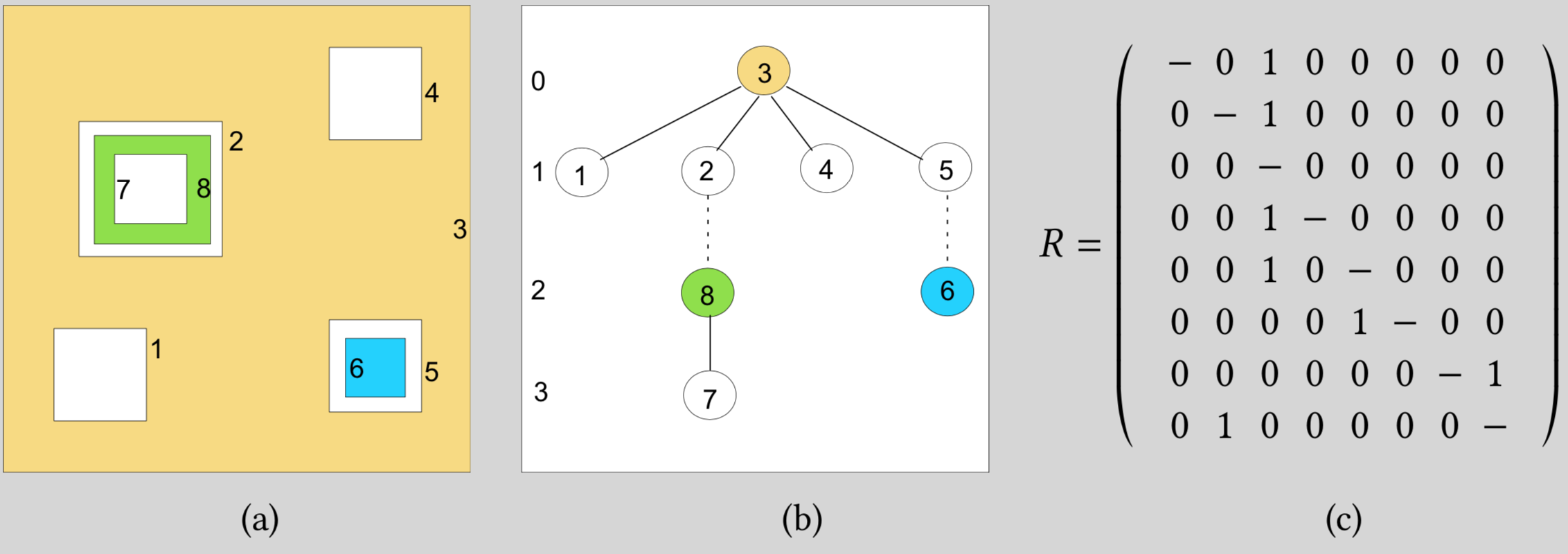} 
   \caption{Non intersecting cycles within a 2D cellular complex with three connected components and only three cells, denoted by the image colors: (a) cellular complex; (b) graph of the \emph{reduced} containment relation $R$ between shells, with dashed arcs of even depth index. Removing the dashed arcs produces a \emph{forest} of small \emph{trees}; (c) matrix of transitively reduced $R$. Note that the ones {equate} the number of edges in the graph. }
   \label{fig:shells}
\end{figure}

\subsubsection{Shell discovering and assembling/removing}\label{shells}

Then add to the set $S$ the empty cycles (shells) already included within some non-contractible cell. We  call them \emph{input holes}. We discover the (unmodified) input holes by direct inspection of matrices $[\partial_3^p]$, since holes are represented there as \emph{pairs of columns with zero sum}. In fact, each input hole, if non intersected by other data, returns unmodified, and produces \emph{two opposite  columns} in the component matrix $[\partial_3^p]$ it belongs to. Each corresponding pair of columns has all non-zero elements on the same rows, but with opposite signs (orientations).
The inspection is done by a sort of scan-line algorithm working on the rows of each $[\partial_3^p]$ matrix, that recognizes the emerging pairs of opposite columns, and stores each pair (candidate hole) until its contents eventually differ. The  algorithm proceeds by moving from a row to the  next one, until the bottom row is reached and the set, possibly empty, of holes is returned. One column from each pair (with proper sign---see~\ref{poset}) is removed from the component matrix $[\partial_3^p]$, and its opposite 2-cycle is added to the set $S$ of shells.

\subsubsection{Transitive reduction of shell poset}\label{poset}

The antisymmetric \emph{containment relation} between 2-cycles in $S$ is computed for $i,j\in [1,\ldots,h], i<j$, by the containment test $\mathit{PointInPolygon}(u_0^i,c^j)$ between a single unit chain $u_0^i \in c^i$ and the cycle $c^j$, with $(c^i,c^j)\in S$. Then, the transitive reduction $R$ is extracted. 
If the edge-set of the $R$ tree is empty, no disjoint component of $X_3$ is contained inside another one, and both $X_3$ and $\partial_3$ may by assembled by disjoint union of 3-cells of $X_3^p$ and columns of $[\partial_3]^p$, $(1\leq p\leq h)$, respectively.
If, conversely,  the above is not true, then reduce $R$ to a forest of small trees by cancelling the arcs at even distance from root (see Figure~\ref{fig:shells}) and, for each arc $(i,j)$, discover which \emph{cell} of the container component $X^i$ actually contains the contained component $X^j$, \emph{i.e.}, its shell $c^j$ and its possibly non-empty interior. More complex intersecting situations are impossible by construction, since the components are a priori disjoint. Therefore, in case of containment, one component is necessarily contained in some \emph{empty cell} of the other.

\subsection{Computational interlude}\label{sec:prologue}

For reader's convenience, we give in Appendix~\ref{sec:appendix} definitions and facts about computing with chains and cochains, and provide some examples of elementary topological computation with chains, cochains and their operators.
Be it noted that our \emph{topology representations}, \emph{i.e.},
\begin{itemize}
\item 1-,2-,3-cells as either 0-chains or p-cycles ($p=1,2$), as \emph{sparse 1-arrays} of signed numbers,
\item 1-,2-,3-complexes as bases of linear spaces of $p$-chains and graded linear transformations, represented as \emph{sparse 2-arrays} of signed numbers,
\end{itemize}
have smaller space complexity than common data structures of well-known efficient  representations~\cite{Woo:85,Dicarlo:2014:TNL:2543138.2543294} of Solid Modeling. For example, with regard to the B-rep of a closed 2-manifold $A$, we have $\mathit{Space(A)} = \mathit{Space([\partial_2])}+\mathit{Space([\partial_1])} = 2\#\texttt{E}+2\#\texttt{E}$, where $[\partial_2], [\partial_1]$ are sparse matrices of boundary operators, and $\#\texttt{E}$ is the number of unit 1-chains (edges). Hence the storage required by $\mathit{Space(A)}$ is equal to $2/3$ of \emph{half-edge}~\cite{Muller:78}, largely used in Computational Geometry, and $1/2$ of \emph{winged-edge}~\cite{Baumgart:1972:WEP:891970}, often used as a reference representation for manifold Solid Modeling~\cite{Ala:1992:PAB:616022.617736}. 

The cardinality of \emph{all} the incidence/adjacency relations between p-cells and q-cells $(1\leq p,q \leq 3)$ is also minimal, according to~\cite{Woo:85}.  For example, $\#\texttt{FV} = O(Space([\partial_2] [\delta_1]) = 2\#\texttt{E}$), where \texttt{V} are the vertices of a complex and \texttt{EV}, \texttt{FV} are binary incidences of edges and faces with vertices. Therefore, every set of local queries about the $3\times 3$ incidences/adjacencies between $p$-cells can be answered by multiplication, via software kernels for \emph{sparse matrix} product and transposition, just by collecting the coordinate vectors of unit chains, ``subject''  of elementary queries, as \emph{columns} of a sparse $Q$ matrix, and by left-multiplying $Q$ times one/two operator matrices $[\partial_1]$ and/or $[\partial_2]$, suitably ordered and/or transposed~\cite{Dicarlo:2014:TNL:2543138.2543294}, to get the algebraic equivalent of multiple database queries at once.

\section{Chain-based arrangement algorithms}
\label{sec:algorithms}

In this section, we provide a slightly simplified pseudocode\footnote{The actual implementation is available as open-sourced Julia package in Github: \href{https://github.com/cvdlab/LinearAlgebraicRepresentation.jl}{ https://github.com/cvdlab/LinearAlgebraicRepresentation.jl}.} of the main algorithms introduced in the previous section,
and discuss their worst-case complexity.  The pseudocode style {is a blend of} Python and Julia styles. 
A few words about  notations:  greek letters are used for the \emph{cells} of a space partition, and roman letters for  \emph{chains} of cells, all coded as either signed integers or sparse arrays of signed integers.  $[\partial_d]$ or $[c]$ stand for general matrices or column matrices, {respectively}, whereas $\partial_d[h,k]$ or $c[\sigma]$ stand for their indexed elements. Also, $|c|$ stands for \emph{unsigned} (nonzero) indices of the (sparse) array $[c]$.  
The \emph{accumulated assignment} statement $A\ $+=$\ B$ stands for $A = A + B$, where the meaning of ``$+$'' symbol depends on the {context}, \emph{e.g.} may stand either  for sum (of chains), or for union (of sets), or for concatenation (of matrix columns). Analogously, $A\ $-=$\ B$ stands for $A = A - B$.

\subsection{Arrangements of 2-cells (Merge algorithm)}\label{arrangement-of-2-cells} 

The sequence of computations performed on each 2-cell $\sigma\in\mathcal{S}_2\subseteq \mathcal{S}$ is discussed in the following. A visualization of the decomposition process, discussed in Example~\ref{ex:square}, is shown in Figure~\ref{fig:abutting}. 

\subsubsection{Fragmentation of a 2-cell}\label{fragmentation}

In a first stage, the subset $\mathcal{I}(\sigma)$ of 2-cells of \emph{potential intersection} with $\sigma$ is computed (see Figure~\ref{fig:process}c), by intersecting the results of three queries upon 1D \emph{interval trees}, generated at the beginning of the pipeline. 
Each 1D interval-tree is based on one dimension of the 3D containment boxes of input 2-cells. Then, the set $\Sigma = \{\sigma\}\cup\mathcal{I}(\sigma)$ is transformed so that $\sigma$ is mapped into the $z=0$ subspace (Figure~\ref{fig:process}d).  
In $\E^3$ the mapped 2-cells in $\Sigma\cap\{(x,y,z)|z=0\}$ are used to compute a \emph{set of line segments} in $\E^2$, generated by intersection of edges of 2-cells in $\Sigma$ with the 2D plane, and by join (convex combination) of alternate pairs\footnote{The 2-cell being intersected with $z=0$ may be non-convex, and its 1-cells may produce and even number $k>2$ of intersection points along the same line.} of intersection points of boundary edges, along the intersection line of each 2-cell in $\Sigma$ (Figures~\ref{fig:process}e and~\ref{fig:2D}a). 

\begin{algorithm}[h]
\SetAlgoNoLine
\DontPrintSemicolon
\KwIn{$\mathcal{S}_2\subseteq \mathcal{S}_{d-1}$ $\qquad$ { \emph{collection} of all 2-cells from $\mathcal{S}_{d-1}$ input in $\E^{d}$}}
\KwOut{$[\partial_2]$ $\qquad$ {signed CSC (Compressed Sparse Column)  matrix}}
$\widetilde{\mathcal{S}_2} = \emptyset$ $\qquad$ {{ initialisation of collection of local fragments }}\;
\For($\qquad$ {{ for each 2-cell $\sigma$ in the input set}}){$\sigma \in \mathcal{S}_2$}{
	$M = \mathit{SubManifoldMap}(\sigma)$ $\qquad$ {{ affine transform \emph{s.t.} $\sigma\mapsto x_3=0$ subspace}}\;
    $\Sigma = M\,(\mathcal{I}(\sigma)\cup\{\sigma\})$ $\qquad$ {{ apply the map $M$ to (possible) incidencies to $\sigma$}}\;
	$\mathcal{S}_1(\sigma) = \emptyset$ $\qquad${{ collection  of line segments in $x_3=0$}}\;
    \For($\qquad$ {{ for each 2-cell $\tau$ in $\Sigma$}}){$\tau \in \Sigma$}{
        $\mathcal{P}(\tau),\mathcal{L}(\tau) = \emptyset,\emptyset$ $\qquad$ {{ intersection points and segment(s) with $x_3=0$}}\;
        \For($\qquad$ {{ for each 1-cell $\lambda$ in $X_1(\tau)$}}){$\lambda \in X_1(\tau)$}{
        	\lIf($\quad$ {{ save intersection of $\lambda$ and $x_3=0$}}){$\lambda\not\subset\{\p{q}\,|\,x_3(\p{q})=0\}$}{
				$\mathcal{P}(\tau)\ $+=$\ \{\p{p}\}$}
		}
		$\mathcal{L}(\tau) = \mathit{Points2Segments}(\mathcal{P}(\tau))$ $\quad$ {{ set of collinear intersections}}\;
		$\mathcal{S}_1(\sigma)\ $+=$\ \mathcal{L}(\tau)$ $\quad$ {{ accumulate intersection segments generated by $\tau$}}\;
	}
	$X_2(\sigma) = \mathcal{A}(\mathcal{S}_1(\sigma))$ $\quad$ {{ arrangement of $\sigma$ induced by a set of 1-complexes}} \;
	$\widetilde{\mathcal{S}_2}\ $+=$\ M^{-1}\,X_2$  $\qquad$ {{ accumulate local fragments, back transformed in $\E^d$ }}\;
}
$[\partial_1] = \mathit{QuotientBases}(\widetilde{\mathcal{S}_2})$  $\qquad$ {{ identification of 0- and 1-cells using $kd$-trees  }}\;
$[\partial_2] = \mathit{TGW}([\partial_1])$ $\qquad$ {{ output computation via TGW algorithm in 2D}}\;
\Return{$[\partial_2]$}
\caption{ \emph{{Subdivision} of 2-cells }}
\label{alg:two}
\end{algorithm}

The {planar processing of the 2-cell $\sigma$} continues by  pairwise executing the line segment intersection, producing a linear graph, as shown in Figure~\ref{fig:2D}b. The dangling~\footnote{In a $d$-complex, \emph{dangling edges} are $p$-cells, p<d, that are not contained in any boundary cycle of a $d$-cell. They are the interior structures of SGC cells. In Solid Modeling terminology, they are called non-regular subsets, whence the term \emph{regularized} Boolean operation.} edges are removed, by computing the maximal 2-vertex-connected subgraphs\footnote{A connected graph G is \emph{2-vertex-connected} if it has at least three vertices and no articulation points. 
A vertex is an \emph{articulation point} if its removal increases the number of connected components of G.}~\cite{vialar2016handbook}, with the Hopcroft's and Tarjan's algorithm~\cite{Hopcroft:1973:AEA:362248.362272}. 
Only the non-external biconnected components enter the following computations, since the other graph parts are either external to $\sigma$ or certainly dangling (1-connected subgraphs), and will  contribute separately to the space arrangement (Figure~\ref{fig:2D}b).  
{Finally}, the oriented 2-chain of the partition $\mathcal{A}(\Sigma)$ is computed  as shown in Figures~\ref{fig:process}g and ~\ref{fig:2D}d, using the TGW in 2D, so generating the $\partial_2(\sigma)$ matrix from $X(\sigma)$.
The fragmentation process is repeated for each $\sigma\in \mathcal{S}_2$, with each geometric map $\mu(\sigma): X_0(\sigma)\to\E^2$ composed with its inverse transformation back to $\E^3$.

\subsubsection{Complexity of 2-cells {subdivision}}

The time complexity of Algorithm~\ref{alg:two} is bounded by the number $n$ of 2-cells in the input collection $\mathcal{S}_{d-1}$ times the worst-case cost required by the {subdivision} of one of them. In turn, this cost depends on the size of the actual input, \emph{i.e.}, on the number of potentially intersecting 2-cells in $ \mathcal{I}(\sigma)$. In all the regular cases we usually meet in computer graphics, CAD meshes and engineering applications, the number of 2-cells incident in (even on the boundaries of) a given cell is bounded by a constant number $k_1$. If the maximum number of 1-cells on the boundary of a 2-cell is $k_2$, then the whole computation of  Algorithm~\ref{alg:two} requires time $O(k_1 k_2 n + A)$, where  $A$ is the time needed to compute the quotient sets, \emph{i.e.},~to glue all  $X_2(\sigma)$ in $\E^d$ space.  When $d=3$, the affine transformations $\T{Q}_\sigma$ of each set $\Sigma$ (see~\ref{facet-arrangements}) are computable in $O(1)$ time; building a static $kd$-tree generated by $m$ points requires $O(m \log^2 m)$; and each query for finding the nearest neighbor in a balanced $kd$-tree requires $O(\log m)$ time on average. The number of occurrences of the same vertex on incident 2-cells is certainly bounded by a small constant $k_3$, approximately equal to $m/v$, where $v=\#X_0$ is the number of 0-cells after the identification processing. The transformation of LAR output to canonical form (sorted 1-array of integers)  is done in $O(1)$ for each edge, so giving $A = O(m \log^2 m) + O(m\log m) + O(1) = O(m \log m)$. In conclusion, {the worst-case running time of Algorithm~\ref{alg:two} is}  $O(k_1 k_2 n + m \log m)$.

\subsection{Quotient sets computation (Congruence algorithm)}\label{quotients}

Small sparse matrices of signed operators $\partial_2(\sigma): C_2(\sigma)\to C_1(\sigma)$ have  already been assembled \emph{independently} in 2D for each fragmented $\sigma$, \emph{i.e.}, for each $X_2^\sigma$, as detailed in the previous Section~\ref{fragmentation}. The output of that pipeline stage is a collection  $\varmathbb{C} := \{ C_\bullet(\sigma), \sigma\in \mathcal{S}_2\}$ of small chain complexes, one for each input 2-cell, embedded in $\E^3$. 
They were buil by repeatedly applying in 2D the TGW algorithm (see Section~\ref{partial_3-computation}) and mapping back the results in 3D.
The quotients of chain spaces modulo the $p$-congruence relations are calculated at this point, starting from $p=0$. Therefore, the unit 0-chains are identified numerically via their geometric maps and snap rounded by numerical identification of nearby-coincident points using a $kd$-tree.  The congruent unit 1-chains and 2-chain are identified symbolically, making use of their {unique} canonical indexed representation. The canonical representation of a unit $d$-chain is the array of sorted indices of the unit elements of its $(d-1)$-cycle.  The 2-cells of the output 2-complex $X_2(\mathcal{S}_2)$ embedded in $\E^3$, written as 1-cycles, \emph{i.e.},~as linear combinations of signed 1-cells, are stored by column in the matrix of the operator $\partial_2: C_2\to C_1$. A 1-cell $\tau$, is written\footnote{As a 0-chain of signed 0-cells in the matrix representation of $\partial_1: C_1\to C_0$.} by convention as $1u_0^{i_k} -1u_0^{i_h}$ {when $k>h$},  and is oriented from $u_0^{i_h}$ to $u_0^{i_k}$.  The conventional rules used in this paper about sign and orientation of cells are summarized at the end of Section~\ref{sec:thechains}.

\begin{algorithm}[h]
\SetAlgoNoLine
\DontPrintSemicolon
\tcc{{\rm\textbf{Pre-condition:} $d$ equals the space $\E^d$ dimension, such that~$(d-1)$-cells are shared by \emph{two} $d$-cells}}
\tcc{{\rm ~}}
\KwIn{$[\partial_{d-1}]\qquad~$ { signed CSC matrix} $(a_{ij})$, where $a_{ij}\in\{ -1,0,1 \}$}
\KwOut{$[\partial_d^+]\qquad~$ { signed CSC matrix of $(d-1)$-cycles}}
$[\partial_d^+] = []\ $; $m,n = [\partial_{d-1}].shape\ $; $marks = Zeros(n)\ $ $\qquad~$ {{ initializations}}\; 
\While{\emph{Sum}$(marks) < 2n$}{
	$\sigma = Choose(marks)$ $\qquad$ {{ select the $(d-1)$-cell seed of column extraction\;}}
	\lIf{$marks[\sigma] == 0$}{$[c_{d-1}] = [\sigma] $}
	\lElseIf{$marks[\sigma] == 1$}{$ [c_{d-1}] = [-\sigma]$}
	$[c_{d-2}] = [\partial_{d-1}]\,[c_{d-1}]$ $\qquad$ {{ compute boundary $c_{d-2}$ of seed cell\;}}
	\While($\qquad$ {{ loop until boundary becomes empty}}) {$[c_{d-2}] \not= []$}{
		$corolla = []$\;
		\For($\qquad$ {{ for each ``hinge'' $\tau$ cell}}){$\tau \in c_{d-2}$}{
			$[b_{d-1}] = [\tau]^t [\partial_{d-1}]$  $\qquad$ {{ compute the $\tau$ coboundary}}\;
			$pivot = \{|b_{d-1}|\} \cap \{|c_{d-1}|\}$  $\qquad$ {{ compute the $\tau$ support}}\;
			\lIf($\qquad$ {{ compute the new adj cell}}){$\tau > 0$}{$adj = Next(pivot, Ord(b_{d-1}))$}
			\lElseIf{$\tau < 0$}{$adj = Prev(pivot, Ord(b_{d-1}))$}
			\lIf($\qquad$ {{ orient adj}}){$\partial_{d-1}[\tau,adj] \not= \partial_{d-1}[\tau,pivot]$}{$corolla[adj] = c_{d-1}[pivot]$}
			\lElse{$corolla[adj] = -(c_{d-1}[pivot])$}
		}
		$[c_{d-1}]\, $+$= corolla$  $\qquad$ {{ insert $corolla$ cells in current $c_{d-1}$\;}}
		$[c_{d-2}] = [\partial_{d-1}]\,[c_{d-1}]$  $\qquad$ {{ compute again the boundary of $c_{d-1}$\;}}
	}
	\lFor($\qquad$ {{ update the counters of used cells}}) {$\sigma\in c_{d-1}$}{$marks[\sigma]\ $+=$\ 1$}
	[$\partial_d^+]\ $+$= [c_{d-1}]$  $\qquad$ { append a new column to $[\partial_d^+]$ }
}
\Return{$[\partial_d^+]$}
\caption{ \emph{Computation of signed $[\partial_d^+]$ matrix} }
\label{alg:one}
\end{algorithm}

\subsection{\texorpdfstring{Computation of \(\partial_2\) and \(\partial_3\)  (TGW Algorithm)}{\textbackslash{}delta\_1 computation (2 pages)}}\label{partial_3-computation}

\subsubsection{Topological Gift Wrapping}

The algorithm was introduced in Section~\ref{tgw-algorithm}. Here we provide a readable pseudocode, with the only \emph{caveat} that it actually computes a redundant set of generators for $C_3$ (resp.~$C_2$),  as minimal connected $2$-cycles (resp.~$1$-cycles) from a $\partial_2$ (resp.~$\partial_1$)  matrix.
A step-by-step formalized example of computation of a unit 2-chain as 1-cycle, using the TGW algoritm, is discussed in Example~\ref{example1}.
The given pseudocode makes use of math symbols and high-level math operations;  its actual implementation in scientific languages like Python or Julia uses sparse arrays and discrete coordinates in $\{-1,0,1\}$, to achieve an efficient execution in terms of storage space and computation time.

Note the precondition of Algorithm~\ref{alg:one}, warning that the method used will compute the $\partial_d$ matrix only for a cell decomposition of $d$-space. In fact, only in this case the $(d-1)$-cells are shared by exactly \emph{two} $d$-cells, including the outer cell. 
This condition implies that the input cellular complex it applies to should be a (possibly non-connected) CW-complex, with all cells homeomorphic to spheres. Note also that the matrix of the boundary operator for the $d$-chain space of a cellular complex with holes as well as inner and outer components will be built starting from output of Algorithm~\ref{alg:one} in a later stage. The termination predicate of Algorithm~\ref{alg:one} is a consequence of the above  property: the algorithm terminates when all incidence numbers in the $marks$ array are 2, so that their sum is exactly $2n$, where $n$ is the number of $(d-1)$-cells, equal to the number of columns in the input matrix $[\partial_{d-1}]$.

\subsubsection{Valid input and unique output}\label{sec:limitations}

{The algorithm works properly with legitimate input. In particular, input $(d-1)$-skeletons must be regular, \emph{i.e.}, without dangling parts, so that every $(d-1)$-cell belongs at most to \emph{two} $(d-1)$-cycles. In 2D, this fact is guaranteed by applying the algorithm separately to each 2-maximally connected component of the 1-skeleton, considered as a graph, and then by merging the results (clearly disconnected). Analogously, in 3D, the adjacency graph of 2-cells should not contain dangling subgraphs. } 

The validity set of the input may contain 2-skeletons of 3-complexes, boundaries of solid models, sets of manifold boundary components of non-manifold solid models. Of course, to apply the algorithm to data which do not determine a partition of the embedding space does not make sense and produces an empty result.  We call them \emph{illegal data}.
When applied to valid input, as described above, the TGW algorithm produces valid output, because always produces a set of generators for $C_d$ that satisfies the Eq.~\ref{eq:valid} below:
\begin{equation}
[\partial_d] = (a_{ij})\quad \mbox{where}\quad 
\sum_{i=1}^{\#X_{d-1}} \sum_{j=1}^{\#X_{d}} |a_{ij}|= 2\,(\#X_{d-1}),
\label{eq:valid}
\end{equation}
The results are also \emph{unique}, modulo reordering, since otherwise two different bases for the linear space $C_d$ would produce two boundary operators that, applied to the total 3-chain (vector of all ones) would return the same boundary cycle, which is impossible.
There are no ambiguities in the algorithm, since in a $d$-complex every two $d$-cells share at most two $(d-1)$-cells, or exactly two if the outer $d$-cell is considered. Also, it halts when this last condition is exactly reached.  Note that the suitable choice of the next ``petals'' from  ``corolla'' (see the pseudocode in Algorithm~\ref{alg:one}) implies that a 2-cell cannot be used more that twice.

\subsubsection{Complexity of 3-cell extraction}
\label{sec:complexity1}

In three dimensions, Algorithm~\ref{alg:one} constructs iteratively (outer \textbf{while}) one unit 3-chain (represented as a 2-cycle, \emph{i.e.}, as a closed 2-chain), building the corresponding column of the matrix $[\partial_3^+]$, and so adding one outer boundary column for each connected component of the input complex, as detailed in \ref{extension}. 


The space complexity of a 3-cell is measured by a set of triples (row, column, value) for each non-zero values\footnote{{The coordinate (\texttt{COO}) representation of sparse matrices~\cite{gemmexp} is an array of triples $(i,j,value)$.}} in its column, \emph{i.e.},~with its representation as cycle of unit 2-chains. Hence, the total number of triples, \emph{i.e.},~the space complexity of the \texttt{COO} representation of $[\partial_3^+]$, is exactly $2n$, where $n$ is the number of 2-cells in the $X_2$ skeleton. 

The construction of a single 3-cell requires the search of the adjacent $\mathit{adj}$ 2-cell for each $\mathit{pivot}$ unit 2-chain in the boundary shell. The search for $\mathit{next}$ or $\mathit{prev}$ 2-cell as $\mathit{adj}$ for each $\mathit{pivot}$ requires the circular sorting of this permutation subgroup of 2-cells incident to each 1-cell on each boundary of an incomplete 2-cycle. Consequently, we have several sorts of small sets, where each set is normally bounded by a very small integer, hence each sort is $O(1)$ timewise, and their total number is upper bounded by the number of $(d-1)$-cells on the $d$-cell boundary (equal to 6 for cubical 3-complexes, and to 4 for simplicial 3-complexes, and a small integer in general). 

The subsets to be sorted are encoded in the columns of the incidence matrix from 2-cells to 1-cells, \emph{i.e.}, by the $i,j$ indices of  non-zero elements of $[\partial_2]$. The computation of the (unsigned) $[\partial_2]$ may be performed through SpGEMM\footnote{SpGEMM is a subroutine for matrix multiplication between two general sparse matrices~\cite{gemmexp}, \emph{i.e.}, no banded, nor Hermitian, \emph{etc.} Name derived from BLAS rules~\cite{Ballard:2015}.} multiplication of two sparse matrices (see~\cite{Dicarlo:2014:TNL:2543138.2543294}), hence in time linear with the size of the output, \emph{i.e.}, with the number of non-zero elements of the   $[\partial_2]$ matrix.  Summing up, if $n$  is the number of $d$-cells and $m$ is the number of $(d-1)$-cells, the time complexity of this algorithm is $O(n m\log m)$ in the worst case of unbounded complexity of $d$-cells, and roughly $O(n k\log k)$ if their $(d-1)$-cycle complexity is bounded by $k$.

\subsection{Isolated shells (Holes algorithm)}\label{extension}

In the general case, both the outer cell and the inner cells may contain holes and/or isolated components. Talking of isolated holes is improper, since holes  are not empty within an arrangement, \emph{i.e.}, a partition of the ambient space, but contain an isolated component within their boundary represented as a ($d-1$)-cycle. The aim of this section is to discuss the handling of isolated components and their boundaries, to be considered holes within their container cells.

The TGW Algorithm~\ref{alg:one} produces CW-complexes, despite the fact that the subdivision of Merge Algorithm~\ref{alg:two} may generate non contractible $d$-cells, \emph{i.e.}, cells with holes {(but without isolated points). These spaces are handled by combining standard CW-complexes, \emph{i.e.}, with cells homeomorphic to spheres, and by adding $d$-cells to the interior of $d$-cells. }
In other words, the boundary of holes in a cell coming from disconnected sub-complexes is merged in the container cell. The orientation is handled depending on the parity of relative containment relation. The management of isolated boundaries concerns essentially the adjoining/removal of columns in the final boundary matrix.

\subsubsection{Synthesis of the whole pipeline}\label{preview-tgw}

We need to consider two main issues: (a) the computation of {maximal} connected components of $X_{d-1}$ may produce $h>1$ {disconnected} \emph{$d$-components} of the output complex $X_d$; (b) the inclusion of components within single cells of the output $d$-complex: see Figure~\ref{fig:shells}.  In the following we list the main stages of  the \emph{Holes} algorithm to take care of these issues. Our goal is the computation of both the $X_d$ skeleton, and the $\partial_d$ operator for spaces with multiple components nested into holes.  Note that exactly the same points apply (scaled in dimension) before and after TGW execution, for both \emph{local} arrangements in 2D and the \emph{global} arrangement in 3D, respectively. This fact gives a cue for a possible multidimensional extension.

\subsubsection{\texorpdfstring{Non-intersecting shells}{\textbackslash{}delta\_2 computation (2 pages)}}\label{delta_2-computation}

If the shell-set $S \not= \emptyset$, then the $h$ \emph{isolated} boundary components $(0\leq p\leq h)$ in $S$ must be compared with each other, to determine their relative containment, if any, and consequently their orientation. 
The $h\times h$ binary and antisymmetric matrix $M=(m_{ij})$ of the relation is built, by computing each element $m_{ij}$ $(i<j)$, with a single point-cycle ray firing, because the two corresponding cycles (columns $i$ and $j$) are guaranteed not to intersect. 
The attribute of $c_j$ as outer/inner, and hence its relative orientation is given by the parity of $c^j$ in $R$.
When the cancellation \ref{preview-tgw}.4 of empty cells has been performed for all ``solid'' arcs of $R$ (see Figure~\ref{fig:shells}), the updated matrices $[\partial_d]^p$ can be assembled into the final $\partial_d$ operator matrix.

\begin{algorithm}[htbp]
\SetAlgoNoLine
\DontPrintSemicolon
\KwIn{\texttt{LAR}${}_{d-1}$, $[\partial_{d-1}]\qquad\qquad~$ { for $d=3$: \texttt{FV}, $\partial_2$}}
\KwOut{\texttt{LAR}${}_{d}$, $[\partial_d]\qquad\qquad~$ { for $d=3$: \texttt{CV}, $\partial_3$}}
$N = \Lambda_2 \cup \Lambda_0$; $\quad A \subseteq \Lambda_2\times\Lambda_0$; $\quad G = (N,A)$ $\qquad\qquad~$ {{ initializations}}\; 
$\mathcal{G} = \{ G^p\ |\ 1\leq p\leq h \} \leftarrow \mathit{ConnectedComponents}(G)$  $\qquad\qquad~$ {{ partition of $G$ into $h$ connected components}}\; 
$\mathcal{X}_{d-1} = \{ (X_{d-1}^p, \partial_{d-1}^p)\ |\ 1\leq p\leq h \} \leftarrow \mathit{Rearrange}(\mathcal{G})$  {$\qquad\qquad~$ {{ partition of $X_{d-1}$ into $h$ connected components}}}\;
$S = []$ $\qquad\qquad~$ {{ initialize the sparse array of \emph{shells}}}\; 
\For({$\qquad${ for each connected component of $(d-1)$-skeleton}}){$p \in \{ 1,\ldots, h \}$}{
	$[\partial_d^+]^p = \mathit{Algorithm\_1}([\partial_{d-1}]^p)$ $\qquad$ {{ compute the minimal $d$-cycles of a component of complex}}\;
	$(c^p, \partial_d^p) = \mathit{Split}([\partial_{d}^+]^p)$ $\quad$ {{ split the component into the exterior $(d-1)$-cycle and the boundary $\partial_d^p$}}\;
	$S\ $+=$\ [c]^p$ $\qquad\qquad~$ {{ append the boundary shell to the shell array}}\; 
}
\For({$\qquad${ for each shell pair $(c^i,c^j)\in R$,\ }}){$i,j \in \{ 1,\ldots, h \},i<j$}{
	$(R[i,j],R[j,i])::\mathit{Bool}\times\mathit{Bool} \leftarrow \mathit{PointSet}(u_0^i\in c^i,c^j)$ {$\qquad${ containment test of $u_0^i$ in $c^j$ }}\;
}
$R = \{ (i,j) \} \leftarrow \mathit{Tree(TransitiveReduction}(R\,))$ {$\qquad${ set of arcs of reduced containment tree of shells }}\;
\If({$\qquad${ if the containment tree of shells is not empty}}){$R \not=\emptyset$}{
	\For({$\qquad${ for each shell pair $(c^i,c^j)$ such that $dist(c^j)\%2\ $!=$\ 0$}}){$(i,j) \in R$}{
		$\rho = \mathit{FindContainerCell}(u_0^i,c^j,\texttt{LAR}_{d-1})$ {$\qquad${ look for a $d$-cell $\rho$ such that $u_0^i\in  |c^i| \subseteq |\rho|\subseteq |c^j|$ }}\;
		$[\partial_d]^j\ $-=$\ \partial_d^j[\rho]$ {$\qquad${ remove $\rho$ from $\partial_d^j$ }}\;
	}
	}
$\partial_d = [\partial_d^1 \cdots \partial_d^p \cdots \partial_d^h]$ {$\qquad${ return the aggregate $\partial_d$ operator }} \;
$\texttt{LAR}_{d} = [\cup_k \texttt{LAR}_{d-1}(c^k = \partial_d[\cdot,k]),\ \mathit{for\ } k\in\mathit{Range}(Cols(\partial_d))]\qquad$ {{ for $d=3$: \texttt{LAR}${}_{d} = \texttt{CV}$}}\;
\Return \texttt{LAR}${}_{d}$, $[\partial_d]$

\caption{ \emph{Non-intersecting shells}}
\label{alg:four}
\end{algorithm}

\subsubsection{Complexity of shell management}
\label{sec:inclusion}

The computation of the connected components of a graph $G$ can be performed in linear time~\cite{Hopcroft:1973:AEA:362248.362272}. The recognition of the $h$ shells requires the computation of $[\partial_{d}^+]^p$ ($1\leq p\leq h$) and the extraction of the boundary of each connected component $X_{d}^p$ . To compute the reduced relation $R$ we execute $O(h^2)$ point-cycle containment tests, linear in the size of a cycle, so spending a time $O(h^2 nr)$, with $h$ number of shells, and $n$ average size a cycle. Actually, the point-cycle containment test can be easily computed in parallel, with a minimal transmission overhead of the arguments.
The restructuring of boundary submatrices has the same cost of the read/rewrite of columns of a sparse matrix, depending on the number of non-zeros of $[\partial_3]$, and hence is $O(n \#X_d)$, \emph{i.e.}, linear with the product of the number of $d$-cells and their average size $n$ as chains of $(d-1)$-cells, with $n$ size of the average isolated cycle.

\subsection{The whole picture}\label{whole-picture}
A short synthesis of sequential steps of the whole computational pipeline, from input collection to chain complex output, follows in the more general setting, with both isolated components (within the outer cell), and possibly nested isolated components (within holes in inner cells).
\begin{description}
    \item[Input]  Facet selection, \emph{i.e.}, construction of the collection $\mathcal{S}_{d-1}$ from $\mathcal{S}_d$, using \texttt{LAR}.
    \item[Indexing]	Spatial index made by intersection of  $d$ interval-trees on bounding boxes of $\sigma\in\mathcal{S}_{d-1}$.  
    \item[Decomposition]  Pairwise $z=0$ intersection of line segments in $\sigma\cup\mathcal{I}(\sigma)$, for each  $\sigma\in\mathcal{S}_{d-1}$.
    \item[Congruence]	Graded bases of equivalence classes $C_k(U_k)$, with $U_k=X_k/R_k$ for $0\leq k\leq 2$.
    \item[Connection]	Extraction of $(X_{d-1}^p,\partial_{d-1}^p)$,  maximal connected components  of $X_{d-1}$ ($0\leq p\leq h$).
    \item[Bases]	Computation of redundant cycle basis $[\partial_{d}^+]^p$ for each $p$-component, via TGW.
    \item[Boundaries]	Accumulation into $H +{}\!\!\!= [o]^p$ (hole-set) of outer boundary cycle from each $[\partial_d^+]^p$.
    \item[Containment]	Computation of antisymmetric containment relation $S$  between $[o]^p$ holes in $H$.
    \item[Reduction]	Transitive $R$ reduction of $S$ and generation of forest of flat trees $\langle[o_d]^p, [\partial_{d}]^p\rangle$.
    \item[Adjoining]	of  roots $[o_{d}]^r$ to (unique) outer cell, and non-roots $[\partial_{d}^+]^q$ to container cells. 
    \item[Assembling]	 Quasi-block-diagonal assembly of matrices relatives to isolated components $[\partial_d]^p$.
    \item[Output] Global boundary map $[\partial_d]$ of $\mathcal{A}(\mathcal{S}_{d-1})$, and reconstruction of 0-chains of $d$-cells in $X_d$.
\end{description}

\section{Relevant literature}\label{other-works}

In this section we mention some relevant connections of the present approach with recent papers concerning similar topics, and discuss some remarks in relation to our own work.  

In~\cite{Alayrangues2015} a topological approach to homology is introduced for subclasses of subdivided spaces constructed by combinatorial and generalized maps.  Generalized map  (\emph{Gmap}) is a combinatorial model which allows for representing and handling subdivided objects~\cite{Lienhardt:2014} via  ``connecting darts'' between cell pairs. Gmaps are used to describe the topology of manifold-like cellular objects where $p$-cells are homeomorphic to $p$-spheres. 
Alayrangues, Damiand, Lienhardt, and Peltier give an algorithm to build signed boundary maps for $0\leq i\leq 3$, focusing on the equivalence between computing homology via Gmaps and via simplicial complexes. 

The time complexity of boundary maps in~\cite{Alayrangues2015} is linear in the number of incidence numbers. In~\cite{Dicarlo:2014:TNL:2543138.2543294}, we already obtained the same result, linear in the sparse output size, for computation via SpGEMM multiplication (see, \emph{e.g.},~the Example~\ref{ex:example0}) when the input complex is known. The   complexity of TGW for computing the \emph{unknown} $X_d = \mathcal{A}(\mathcal{S}_{d-1})$ is obviously higher (see Section~\ref{sec:complexity1}), and equates the standard in Solid Modeling. The main difference with our approach is that Alayrangues and colleagues start from a \emph{given} Gmap cellular model, whose construction is quite complex and requires interactive operations with a graphical user interface or a symbolic logic systems (such as INRIA's Coq~\cite{DEHLINGER2014869}) with a formal specification language. If simplicity metric matters, our linear algebraic representation of chains with sparse arrays compares well with chains of Gmaps.

With Selective Geometric Complex~\cite{Rossignac:1991:CNG:115604.115606}, Rossignac and O'Connor proposed a significant extension of topological CW-complexes, allowing for $p$-cells with  structures of dimension $0\leq k<p$ internal to cells, \emph{i.e.}~not necessarily embedded in a  cell boundary. The  selection bit associated to each cell allows selective choice of substructures. The association of incidence and local ordering among incident cells is maintained via hierarchical links, analogous to the Hasse diagram between vertices, edges, and faces, as well with geometric extents of non-linear surface patches.
Two attributes for \emph{c.boundary} and \emph{c.star} of a cell return the cells on its boundary and those it is on the boundary of. The search for boundary of more complex substructures is algorithmic.

In the present paper, we conversely use graded and combinable linear operators for  boundary and coboundary to traverse, both locally and globally, the incidence hierarchy. Hence we obtain, via multiplication of sparse matrices and vectors, a complete linear characterization of the space topology. Even local updates to topology, via Euler operators, can be done algebraically~\cite{ieee-tase}.  Another significant difference concerns the large amount of information and pointers associated by SGC to each cell, including
extent, dimension, boundary, activity bit, and extendable attributes. Conversely, in the present paper an oriented unit $p$-chain is characterized only by a signed integer index to $U_p$ basis, and by its signed $(p-1)$ boundary cycle, stored as a sparse column in $[\partial_{p}]$. Of course, all topological queries, both local and global, are allowed by suitable SpGEMM multiplication. We  allow for nesting inner cycles (holes) and substructures to the cells, but not for explicit containment of internal edges and points.  The most part of topological algorithms is algebraic in nature.

\cite{Zhou:2016:MAS:2897824.2925901}, by Zhou, Grinspun, Zorin, and Jacobson,  computes mesh arrangements for solid geometry, takes as input any number of triangle meshes, resolves triangle intersections in 3D, and assigns a winding number vector to subdivided cells, to evaluate variadic Boolean expressions.
Their data are represented by (small) BSPs enriched with explicit convex surface patches on nodes, and adjacency structure between nodes, together with a large amount of additional information.
``The crux of our method is construction of the mesh arrangement data structure, consisting of cells annotated with winding numbers, patches and their adjacency graph, that allows us to extract results of a variety of operations from the arrangement'' (Zhou et al., page 3). 
Just consider their example where 10$K$ geometric models are intersected. It seems reasonable to assume that each of them weights for at least 1$K$ triangles, producing winding number vectors, a forest of BSP trees, incidence structures, \emph{one-to-one} with cells of the resulting arrangement. Telling how the models are distributed in space would give the reader a clue about  the amount of subdivided 3-cells, and its growth function. 

Then, compare the approach~\cite{Zhou:2016:MAS:2897824.2925901}  with our topological method, where each unit $d$-chain (cell) is described just by the sparse array of signed indices of cells on its boundary cycle, and each input 2-cell produces its \emph{(local) fragmented} chain complex of maps, for a total number of 2D complexes equal to that of input 2-cells before 2D fragmentation.
The approach~\cite{Zhou:2016:MAS:2897824.2925901} to variadic Boolean expressions, while very bright, actually requires a scanning of the full stack of data to extract the result of every Boolean expression. Zhou, Grinspun, Zorin, and Jacobson claim that ``the method is variadic, operating on any number of input meshes.'' Actually, their approach can be split into two stages: first, \emph{adding iteratively meshes} to an arrangement;  second, executing all classic Boolean operations. Contrariwise, we do not add each input to the previous result but, in decomposition stage, operate \emph{independently} on each input 2-cell (see~\ref{quotient-set}), according to an embarrassingly parallel data-driven approach. It is also remarkable that the present approach works with more general meshes: sets of 2-manifolds  with- and/or without-boundary, sets of non-manifolds, sets of 3-manifolds, \emph{etc.}, versus just sets of triangle meshes. We do not discuss it here, but extending  our approach to Boolean operations and to Boolean functions is straightforward.

An extremely fast mesh repairing algorithm with guaranteed topology is described in 
\cite{Attene:2014:DRS:2953208.2953514}, mostly based on floating point arithmetics, and requiring exact arithmetics only in relatively few situations. At variance with Attene~\cite{Attene:2014:DRS:2953208.2953514}, we not discriminate between manifold and non-manifold case, and do not use any special data structure in any steps of the pipeline, except 1D interval-trees and $kd$-trees for acceleration. By identifying the conditions that make floating-point arithmetics not reliable, J.R.~Shewchuk, the author of the Triangle library\footnote{\href{https://www.cs.cmu.edu/~quake/triangle.html}{https://www.cs.cmu.edu/~quake/triangle.html}}~\cite{shewchuk96b,SHEWCHUK200221} used in our method, identifies the key for fast robust geometric predicates in adaptive precision floating-point arithmetic~\cite{Shewchuk:1996:RAF:237218.237337}. He wrote that: ``the techniques Priest and I have developed are simple enough to be coded directly in numerical algorithms, avoiding function call overhead and conversion costs.''  In fact, the numerical results we obtained on triangulations with large arrangements of 2D line segments are very fast. We ported the CDT (Constrained Delaunay Triangulation) functions from his C library to Julia language\footnote{\href{https://github.com/cvdlab/Triangle.jl}{https://github.com/cvdlab/Triangle.jl}}, and used it to triangulate on-the-fly each non-convex 2-cell, in order to correctly compute the ordering of ``corollas'' 2-cells around ``pivot'' 1-cells in the 3D TGW. 

In \cite{Campen:2010}, Campen and Kobbelt present a technique to implement operators that modify the topology of polygonal meshes at intersections and self-intersections, by combining an adaptive octree with nested binary space partitions. An analogous decompositive technique was introduced in the geometric language PLaSM~\cite{Paoluzzi:1995:GPP:212332.212349,Paoluzzi2003a}\footnote{\href{https://github.com/plasm-language/pyplasm}{https://github.com/plasm-language/pyplasm}} since 2004 by Scorzelli, Paoluzzi and Pascucci, in the contest of progressive geometry detailing allowing parallel modeling with BSP trees~\cite{Paoluzzi:2004:PDB:1217875.1217907,ScorzelliPP-PSM2008}. The technique is now being substituted by methods given in the present paper, since it does not guarantee sufficient robustness and speed.
In \cite{Guibas:98}, Guibas and Marimont describe a dynamic algorithm to compute the arrangement of a set of line segments in the \emph{digital plane}, and to snap the intersection points at the pixel centers. At variance with them, we snap small clusters of very close (numerically ``quasi-congruent'') points, rounding at the center of their $\epsilon$-neighboroughs in 2D, with average diameter of $10^{-16}$, close to the resolution of IEEE-754 binary floating point.

Barki, Guennebaud, and Foufou
claim that~\cite{BARKI20151235} presents  an exact, robust, and efficient method to execute regularized Booleans on general 3D meshes. They use a triangulation of all faces, and reduce the intersection of two surfaces to the 3D intersection of two triangles. Their simple decomposition process for intersecting faces is very similar to the old paper~\cite{Paoluzzi:1989:BAO:70248.70249} by Paoluzzi and his students.  Note that both~\cite{BARKI20151235} and~\cite{Paoluzzi:1989:BAO:70248.70249} contain a procedure for computation of regularized Boolean operations including isolated shells. The novel point about this matter here is that  in the present paper outer and inner oriented shells, as well as isolated components, are handled through signed sum of closed chains and implemented as sum or difference of their vector coordinates in $\Z$ or $\Z/3\Z$. 

Finally, we recall that 
\emph{Half-edge}, the smallest known efficient representation~\cite{Muller:78} of topology of  planar graphs and closed 2-manifolds by  Muller and Preparata, largely used in Computational Geometry for triangulations and Voronoi diagrams, as well as in meshes for games, requires $6\#\texttt{E}$ space. With the $[\partial_1]$ \emph{and} $[\delta_1]$ operators given in the present paper, we obtain for this class the optimal size $\Omega(n) = {4\#\texttt{E}}$, equal to the input size: two vertices and two faces per edge. It is well known~\cite{Woo:85} that both \texttt{EV} and \texttt{EF} relations weight for $2\#\texttt{E}$, i.e., equal to the space occupied by $[\partial_1]$ {or} $[\partial_2]$ maps, which also allow for the algebraic equivalent of multiple database queries at once---via product of the operator matrix times a matrix of unit binary columns, corresponding to single elementary queries..

\section{Past development and prospects of this project}\label{sec:projects}

This last section sketches birth and development of  contents discussed in this paper, provides references to past and present uses, outlines possible extensions, and introduces the prospects of research that are opened by our linear algebraic approach.

The first three authors started this project about computing with chains, cycles, cochains, and (co)boundary in a seminar series on novel algebraic methods for physical simulation and optimization of geometric design, during the sabbatical of V.S. in Rome [year 2000]. This project was awarded  the IBM SUR award in 2003. LAR sparse arrays, big geometric data and geometric services were discussed in many meetings in Rome, Paris, Madison, Berkeley, and Berlin. Our conversations produced some papers~\cite{DiCarlo:2009:DPU:1629255.1629273,ieee-tase,Dicarlo:2014:TNL:2543138.2543294} that started a lasting sequence of web and face-to-face discussions, algebraic experiments and software tests, producing in recent years three open-sourced partial implementations in Python and Julia. Partial implementations were used for software-based experiments of user-tracking and interior geo-mapping in LAR-based  Building Information Modeling (BIM), meta-design of a general hospital, and delivery of web services aiming at deconstruction and reuse of buildings~%
\cite{stag.20151290, visigrapp17:cvdlab, SpiniMDCP-WEB3D2016.bib, paoluzziMS:2014}.  

Currently, some of the authors work to materialize a Julia package~\cite{DBLP:journals/corr/abs-1710-07819} for topological and geometric design, already including a preliminary implementation of the algorithms in this paper.
A~full implementation, covering the handling of isolated components and holes, is under development. A~second version will include vectorization on the GPU using native Julia~\cite{Besard:2017,besard:2017a}. Next, our plan is to port the chain-maps pipeline on Nvidia's GPX-1 (A.P.~has recently obtained the hardware in-house).
We are already using the modeling approach introduced here relying on the Julian open-source library, for rapid development of building models from analysis of Italian Cadastre  documents and 3D models from pictures or 3D images scanned by flying of drones.

We hope that the basic structures and algorithms discussed in this paper may also find some  appropriate use when combined with representations for convolutional neural
networks, based as well on tensors and linear algebra, in order to properly combine image understanding and geometric modeling.
In particular, our approach to compute the
chain complex\footnote{Of very general type, with basis cells possibly non convex and multiply connected.} of an unknown space arrangement 
should match well with deep NNs~\cite{Goodfellow:2016:DL:3086952,Boxel:2016:DLT:3019358}. 
Also our first experiments with topological methods in medical imaging~\cite{Paoluzzi-DCFJ2016,ClementiSSPP-CAD16} look promising.

\section*{Acknowledgements}
We would like to thank the anonymous reviewers who carefully read our manuscript and provided us with many useful comments. Alberto Paoluzzi and Antonio DiCarlo gratefully acknowledge the support of Antonio Bottaro CTO of R\&D at Sogei, now CEO at Geoweb, who believed in our work, and the EU project \href{https://www.medtrain3dmodsim.eu}{medtrain3dmodsim}. {Vadim Shapiro is supported in part by National Science Foundation grant CMMI-1344205 and National Institute of Standards and Technology.

\section{Summary of results and conclusion}\label{conclusion}

We have introduced a novel view on topology computation of space arrangements, that may find good use in disparate subdomains of geometric and visual computing, discussed an original computational architecture based on linear topological algebra, and claim that our approach is in tune with current trends towards hybrid hardware and its more advanced software applications. In particular, in this paper we provide a pseudocode implementation of the full computational pipeline: from a collection of virtual geometric objects to the chain complex $(C_p,\partial_p)$ of their partition of space, giving a full characterization of the topology induced by the input. This result is obtained going beyond simplicial complexes, and working with general piecewise-linear topology with non-contractible cells. A full porting of this approach to Julia, ``the fresh approach to technical computing''~\cite{BEKS14}, is ongoing (a preliminary open-source implementation is available); some parts are already parallelized and others are open to be. Among the strong points we cite: the compact representation; the combinable nature of maps, allowing for \emph{multiple} queries about the $3\times 3$ local topology relations, via fast sparse kernels for multiplication and transposition; the independent fragmentation of input cells through cell congruence; and the topological gift wrapping algorithm. Last but not least, the whole approach seems to be extendable to higher dimension.
We believe that a full real-time implementation of our algorithms on GPUs will generate new techniques for image understanding, in particular when inputs come from next-generation 3D cameras, already on the market and going to be installed on self-driving mobile vehicles. Part of this work was developed within the framework of the IEEE standardization of model extraction from medical images~\cite{10.1109/MC.2014.103}.

\appendix
\section{Appendix}\label{sec:appendix}

For readers' convenience, we recall here a few definitions and facts about computing with chains and cochains, mainly from~\cite{ieee-tase} and~\cite{Dicarlo:2014:TNL:2543138.2543294}. Some simple examples of computations conclude this appendix.  We use greek letter for \emph{cells} and roman letters for \emph{chains}, \emph{i.e.}, for signed combinations of cells.  With some abuse of language, cells in $\Lambda_p$ and unit (singleton\footnote{A set having exactly one element.}) chains in $C_p$ are often identified.

\subsection{(Co)chain Complexes}
\label{chain-complexes-1-page-definitions}

\subsubsection{{Cellular complex}}
\label{sec:definitions}

Let $X$ be a topological space, and
$\Lambda(X) = \bigcup\Lambda_p$ ($p \in \{0, 1,\ldots,d\}$) a
partition of $X$, with $\Lambda_p$ a set of {(relatively) open}, connected, and manifold $p$-cells. 
A \emph{CW-structure} on the space $X$ is a filtration
$\emptyset = X_{-1} \subset X_0 \subset X_1 \subset \ldots \subset  X_{d-1} \subset X = \bigcup_p X_p$,
such that, for each $p$, the \emph{skeleton} $X_p$ is homeomorphic
to a space obtained from $X_{p-1}$ by attachment of $p$-cells in
$\Lambda_p = \Lambda_p(X)$~\cite{hatcher:2002}.
A \emph{CW-complex} is a space $X$ endowed with a CW-structure, and is also
called a \emph{cellular complex}. A cellular complex is \emph{finite}
when it contains a finite number of cells. A {\emph{regular}}
$d$-complex is a complex where 
every $p$-cell ($p < d$) is contained in the
boundary of a $d$-cell.
Two $d$-cells are \emph{coherently oriented} when their common $(d-1)$-cells have opposite orientations. A cellular $d$-complex $X$ is \emph{orientable} when its $d$-cells can be coherently oriented. 
The \emph{support space} $|\sigma|$ of a cell $\sigma$ is its compact point-set.

\subsubsection{{Chain groups}}
\label{sec:thechains}
Chains are defined by attaching coefficients to cells. Since one wishes to add chains, one has to pick coefficients from a set endowed with the structure of a commutative group, or stronger. Let $(G,+,0)$ be a nontrivial commutative group, whose identity element is denoted $0$. A $p$-chain of $X$ with coefficients in $G$ is a mapping $c_p : X \to G$ such that, for each $\sigma \in X_p$, reversing a cell orientation changes the sign of the chain value:
$$
c_p(-\sigma) = -c_p(\sigma).
$$
Chain addition is defined by addition of chain values: if $c_{p}^1, c_{p}^2$ are $p$-chains, then $(c_p^1 + c_p^2)(\sigma) = c_{p}^1(\sigma) + c_p^2(\sigma)$, for each $\sigma \in X_p$. The resulting group is denoted $C_p(X;G)$. When clear from the context, the group $G$ is often left implied, writing $C_p(X)$.
Let $\sigma$ be an oriented cell in $X$ and $g \in G$. The \emph{elementary chain} whose value is $g$ on $\sigma$, $-g$ on $-\sigma$ and 0 on any other cell in $X$ is denoted $g\sigma$. Each chain can be written in a unique way as a sum of elementary chains.
Chains are often thought of as attaching orientation and/or multiplicity to cells: if coefficients are taken from the group {$G = (\{-1,0,1\}, +,0) \simeq (\Z/3\Z,+,0)$}, then cells can only be discarded or selected, possibly inverting their orientation (see~\cite{ieee-tase}).
{A \emph{$p$-cycle} is a \emph{closed} $p$-chain, \emph{i.e.}, a $p$-chain without boundary.}
It is useful to select a conventional orientation to orient cells automatically. 0-cells are considered all positive.  Closed $p$-cells can be given a {coherent} (internal) {orientation} in according with the orientation of the first $(p-1)$-cell in their \emph{canonical representation} {sorted on indices of their $(p-1)$-cycles. Finally, a $d$-cell may be oriented as the sign of its oriented volume.

\subsubsection{{Chain spaces}}

To allow not only for chain addition, but also for linear combination of chains, coefficients should be taken from a set endowed with the structure of a field, such as $(\mathbb{F},+,\times,0,1)$, where $0$ and $1\ne0$ denote, respectively, the additive and multiplicative identities. \emph{Unit} chains are elementary chains whose value is $u = 1\sigma$ for some cell $\sigma$. Each chain can be written in a unique way as a linear combination of unit chains $u\in U$, if the outer cell is not taken into account. Hence, the space of $p$-chains $C_p$ is endowed with a standard (or natural) basis, comprised of all the independent unit $p$-chains. In particular, $\#U_d = \#\Lambda_d -1$. Often, with some abuse of notation, one does not distinguish between a $p$-cell and the corresponding unit $p$-chain.

\subsubsection{{Characteristic matrices}}
\label{sec:characteristic}

Given a set $S=\{s_j\}$, the \emph{characteristic function} $\chi_A: S\to\{0,1\}$ takes value 1 for all elements of $A\subseteq S$ and 0 at all elements of $S$ not in $A$. 
We call \emph{characteristic matrix} $M$ of a collection of subsets $A_i\subseteq S$ ($i=1,\ldots,n$) the binary  matrix $M=(m_{ij})$, with $m_{ij} = \chi_{A_i}(s_j)$. {A  matrix $M_p$, whose rows are indexed by unit $p$-chains and columns are indexed by unit $0$-chains, provides a useful representation of a basis for the linear space $C_p$. Permuting (reindexing) either rows or columns provides a different basis.}   While chains are mostly presented as formal sums of cells, in the actual implementation their signed coordinate vectors are used as \emph{sparse} arrays, and in particular as CSC (Compressed Sparse Column) maps : $\N\to\{-1,0,1\}$.

\subsubsection{{Cochain spaces}}
\label{sec:cochains}

          Cochains are dual to {chains}: $p$-cochains map linearly $p$-chains to  the underlying field $\mathbb{F}$. Unit $p$-cochains, that yield $1$ when evaluated on one unit $p$-chain and $0$ when evaluated on all the others, form the standard basis of the space of $p$-cochains $C^p$. The linear spaces $C_p$ and $C^p$, being isomorphic, can be identified with each other in {infinitely many ways}. Different legitimate identifications, while affecting the \emph{metric} properties of the chain-cochain complex, do not change the \emph{topology} of finite complexes.\footnote{The reader interested in the notions of measured and metrized chains is referred to \cite{ieee-tase,DiCarlo:2009:DPU:1629255.1629273}.} Since we shall use only the topological properties of finite chain-cochain complexes defined by piecewise linear cell complexes in Euclidean space, we feel free to chose the simplest possible identification, consisting in identifying each element of the standard  basis of $C_p$ with the corresponding element of the standard basis of $C^p$. In this paper, we take for granted that chains and cochains are identified in this trivial way.

\subsection{Topology computing with chains}
\label{topology-computing-with-chains-3-pages-examples}

\begin{example}\label{example1}

Figure~\ref{fig:2D-full} shows a fragment of a 1-complex $X=X_1$ in $\E_2$, with unit chains $u_0^k\in C_0$ and $u_1^h\in C_1$. Here we compute stepwise the 1-chain representation $c\in C_1$ of the central 2-cell of the unknown complex $X_2 = \mathcal{A}(X_1)$, using the Topological Gift Wrapping Algorithm~\ref{alg:one}. Refer to Figure~\ref{fig:step-by-step}a-e, repeated below, to follow stepwise the extraction of the 2-cell as 1-cycle. 

\begin{figure}[htbp] 
   \centering\vspace{-3mm}
   \includegraphics[width=0.33\textwidth]{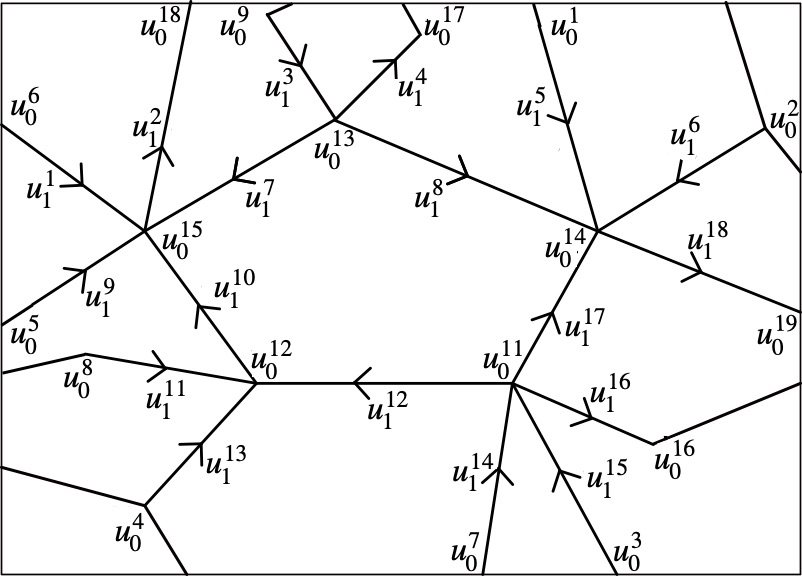} 
   \caption{A portion of the 1-complex used by Example~\ref{example1}, with unit chains $u_0^h \in C_0$ and $u_1^k \in C_1$.}
   \label{fig:2D-full}
\end{figure}

\begin{figure*}[htbp] 
\includegraphics[width=0.2\textwidth]{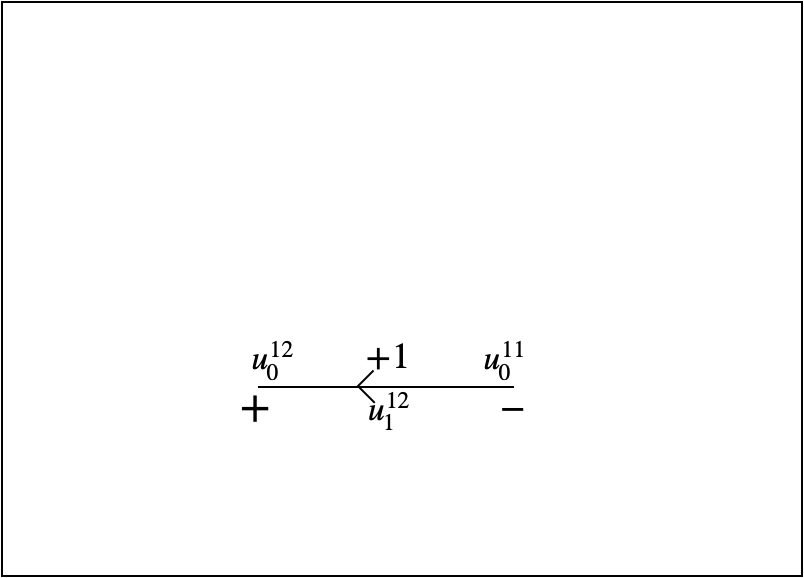}\includegraphics[width=0.2\textwidth]{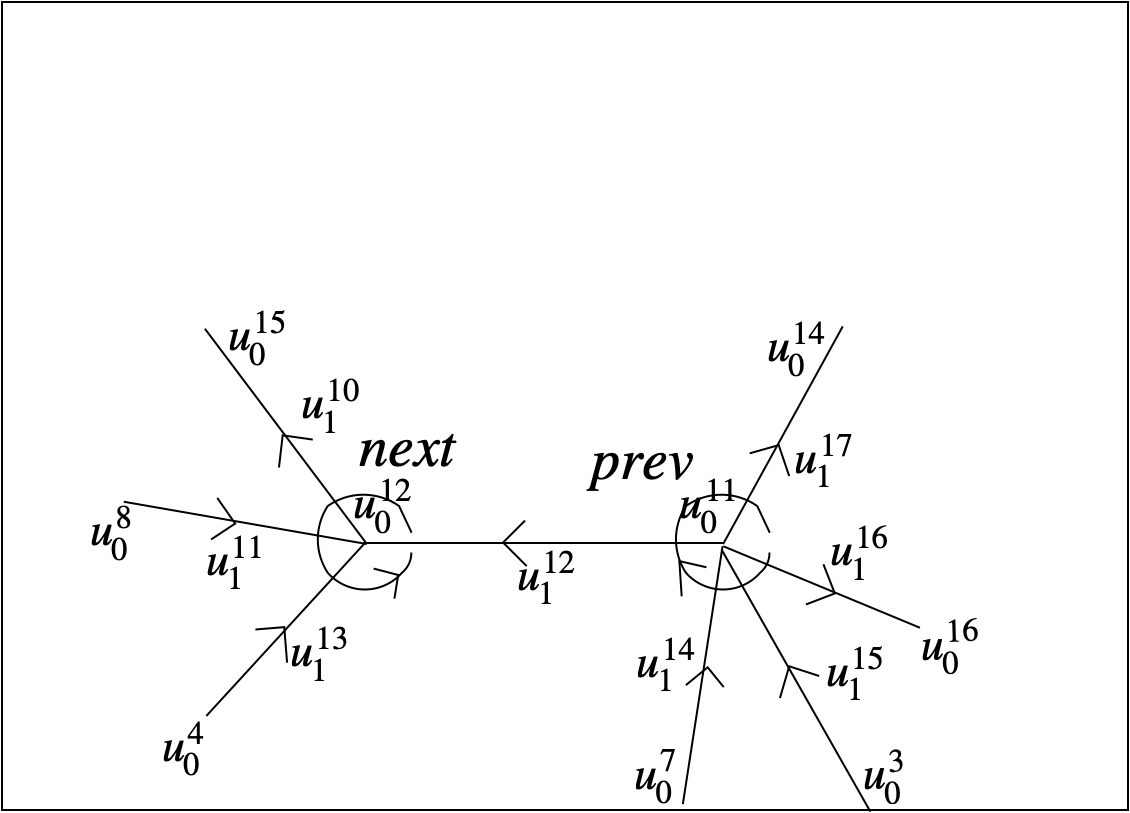}\includegraphics[width=0.2\textwidth]{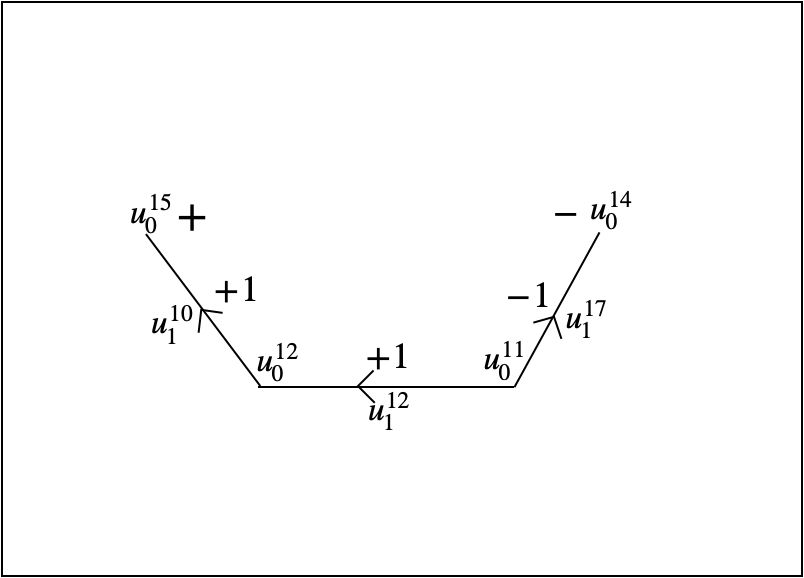}\includegraphics[width=0.2\textwidth]{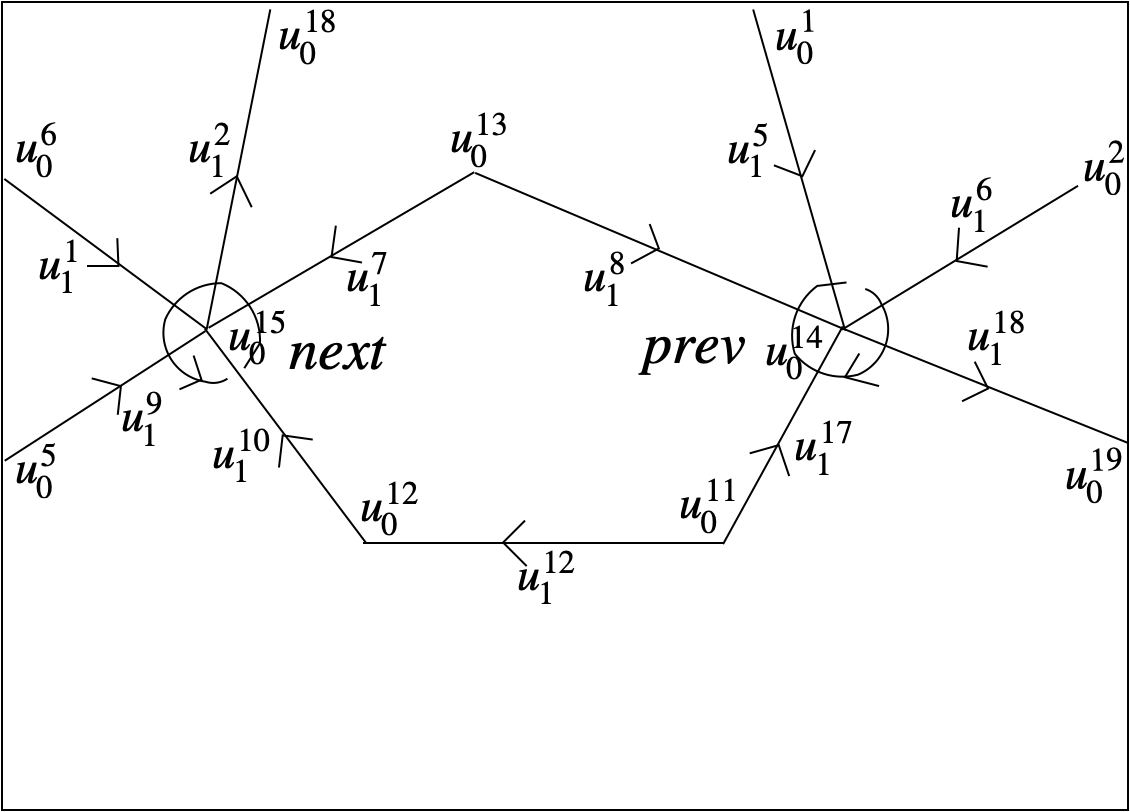}\includegraphics[width=0.2\textwidth]{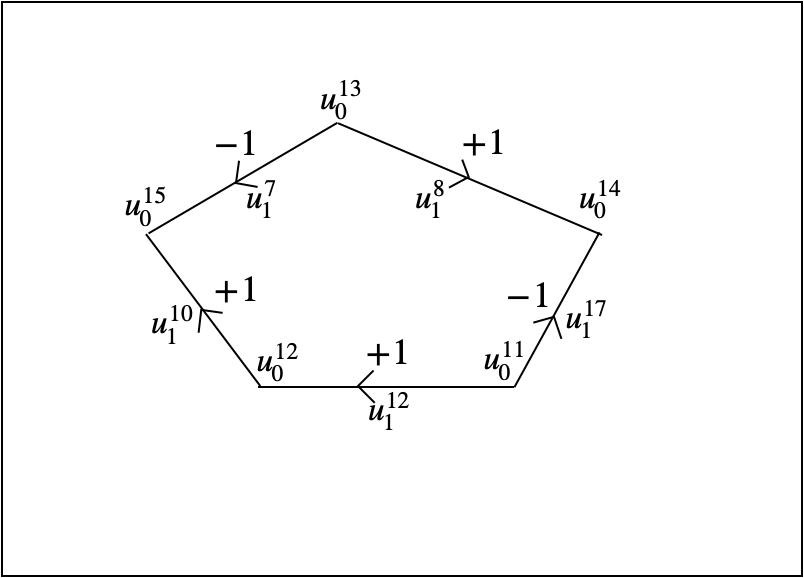}

{\footnotesize\hspace{.09\textwidth}(a)\hfill(b)\hfill(c)\hfill(d)\hfill(e)\hspace{.09\textwidth}}
\end{figure*}

\begin{description}
\item[Step (a)]
Set $c = u_1^{12}$. Then $\partial c=u_0^{12}-u_0^{11}$.
\item[Step (b)]
 $\delta\partial c = \delta u_0^{12}-\delta u_0^{11}$ by linearity. Hence, $\delta\partial c = (u_1^{10}+u_1^{11}+u_1^{12}+u_1^{13})-(+u_1^{12}+u_1^{14}+u_1^{15}+u_1^{16}+u_1^{17})$.
\item[Step (c)]
By computing $\mbox{\emph{corolla}}(c)$, we get
\begin{align*}
\mbox{\emph{corolla}}(c) 
&= c+\mbox{\emph{next}}(c \cap \delta \partial c)\\
&= c+\mbox{\emph{next}}(u_1^{12})(\delta u_0^{12}) - \mbox{\emph{next}}(u_1^{12})(\delta u_0^{11})\\
&= u_1^{12}+\mbox{\emph{next}}(u_1^{12})(\delta u_0^{12}) + \mbox{\emph{prev}}(u_1^{12})(\delta u_0^{11})\\
&= u_1^{12}+u_1^{10}+u_1^{17}.
\end{align*}
If $c$ is coherently orientd, then 
\begin{align}
c = u_1^{10}+u_1^{12}-u_1^{17}, \nonumber\\
\partial c = u_0^{15}-u_0^{12}+u_0^{12}-u_0^{11}+u_0^{11}-u_0^{14} = u_0^{15}-u_0^{14}.\nonumber
\end{align}
\item[Step (d)]
Repeating and orienting coherently the 1-chain yields:
\begin{align*}
\mbox{\emph{corolla}}(c) 
&= c+\mbox{\emph{next}}(c \cap \delta \partial c)\\
&= c+\mbox{\emph{next}}(u_1^{10})(\delta u_0^{15}) - \mbox{\emph{next}}(u_1^{17})(\delta u_0^{14})\\
&= u_1^{10}+u_1^{12}-u_1^{17}+\mbox{\emph{next}}(u_1^{10})(\delta u_0^{15}) + \mbox{\emph{prev}}(u_1^{17})(\delta u_0^{14})\\
&= u_1^{10}+u_1^{12}-u_1^{17}-u_1^{7}+u_1^{8}
\end{align*}
\item[Step (e)]
$\partial\, \mbox{\emph{corolla}}(c) = \emptyset $,
and the extraction algorithm terminates, giving 
\[
c = u_1^{10}+u_1^{12}-u_1^{17}-u_1^{7}+u_1^{8}
\]
as the $C_1(X)$ representation of a basis element of $C_2(X)$, with $X = \mathcal{A}(X_1)$. The coordinate vector of this cycle is therefore accommodated as a new signed column of the yet partially unknown sparse matrix $[\partial_2]$ of the operator $\partial_2: C_2\to C_1$. 
\end{description}

\end{example}

\begin{example}[Chains]\label{examplea}
Unoriented chains take coefficients from $\Z/2\Z = \Z_2=\{0,1\}$. \emph{e.g.}, {a} 0-chain $c\in C_0$ shown in Figure~\ref{fig:ex0-abc}a is given by $c = 1\nu_{1} + 1\nu_{2} + 1\nu_{3} + 1\nu_{5}$. Hence, the coefficients associated to all other cells are zero. So, $[1,1,1,0,1,0]^t$ is the coordinate vector of $c$ with respect to the (ordered) basis $(u_1, u_2, \ldots, u_6)=(1\nu_1, 1\nu_2, \ldots, 1\nu_6)$.
Analogously for the 1-chain $d\in C_1$ and the 2-chain $e\in C_2$, written by dropping the 1 coefficients, as $d = \eta_2 + \eta_3 + \eta_5$ and $e = \gamma_1 + \gamma_3$, with coordinate vectors $[0,1,1,0,1,0,0,0]^t$ and $[1,0,1]^t$, respectively.
\end{example}

\begin{example}[Orientation]\label{examplec}
Figure~\ref{fig:ex0-abc}b shows an oriented version of the cellular complex $\Lambda = \Lambda_0 \cup \Lambda_1 \cup \Lambda_2$, where 1-cells are oriented from the vertex with lesser index to the vertex with greater index, and where {all 2-cells are counterclockwise oriented}. The orientation of each cell may be fixed arbitrarily, since it can always be reversed by the associated coefficient, that is now taken from the set  $\{-1,0,+1\}$.  So, the oriented 1-chain having first vertex $\nu_1$ and last vertex $\nu_5$ is given as $d' = \eta_2 -\eta_3 + \eta_5$, with coordinate vector $[0,1,-1,0,1,0,0,0]^t$.
\end{example}

\begin{example}[Dual cochains]\label{exampleb}
The concept of cochain $\phi^p$ in a space $C^p$ of linear maps from chains $C_p$ to $\R$ allows for the association of a scalar not only to single cells, as done by chains, but also to assemblies of cells. 
{A cochain is hence the association of  discretized subdomains of a cell complex with a global numerical quantity, resulting from a discrete integration over a chain. }
Each cochain $\phi^p \in C^p$ is a linear combination of the {
unit $p$-cochains $\phi_1^p,\ldots, \phi_k^p$}\footnote{{Coincident with $\eta^1_p,\ldots, \eta^k_p$, due to the identification of primal and dual bases.}} (Figure~\ref{fig:ex0-abc}).
The evaluation of a real-valued cochain is denoted as a duality pairing, in order to stress its bilinear property:
\[
\phi^p ( c_p ) = \langle \phi^p, c_p \rangle .
\]
This mapping is orientation-dependent, and linear with respect to ``assemblies of cells", modeled by chains~\cite{hinzl:thesis:2007}. 
\end{example}

\subsubsection{Discrete differential operators}\label{firstexamples}

Let us consider a space partition into cells of dimension 0 ($\nu_i$, $1\leq i\leq 6$), dimension 1 ($\eta_j$, $1\leq j\leq 8$), and dimension 2 ($\gamma_k$, $1\leq k\leq 3$), associated with different additive groups of coefficients. In particular, 
in Figure~\ref{fig:ex0-abc}c the 0-cells are  given arbitrary numbers, \emph{e.g.}, the values of an arbitrary scalar field; whereas the values for 1-cells and 2-cells were computed from these using the co-boundary relations $\delta_0=\partial_1^\top$ and $\delta_1=\partial_2^\top$. Note that they are discrete gradients and curl values associated to 1-cells and 2-cells, respectively. In discrete geometric calculus, we are interested in cochains as functions from chains to reals. The colored numbers on 1- and 2-cells are exactly the evaluation $\phi^k(u_k)=\langle \phi^k, u_k\rangle$ of the dual elementary cochain on each elementary chain.}
In Example~\ref{exampled} we show the numeric values of the matrices.

\begin{figure}[htbp] 
   {\centering
   \includegraphics[width=0.32\linewidth]{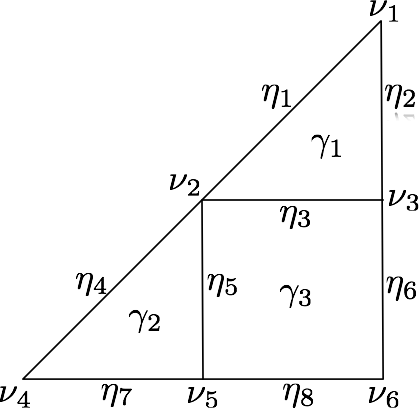}
   \hfill
   \includegraphics[width=0.32\linewidth]{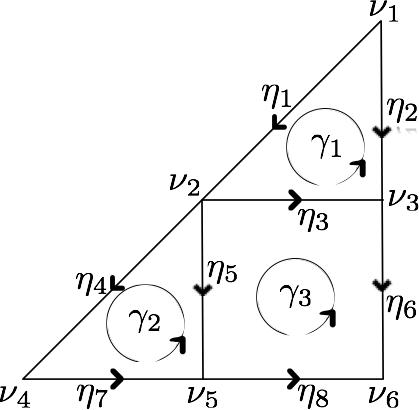}
   \hfill
   \includegraphics[width=0.32\linewidth]{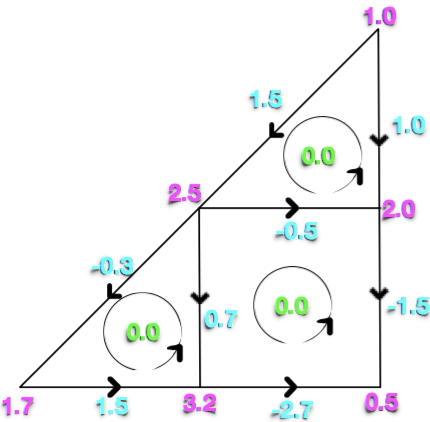}
   
   \small{\hfill(a) \hfill\hfill\hfill (b) \hfill\hfill\hfill (c) \hfill}}
   \vspace{-3mm} \caption{
Cellular complexes with 0-cells {in} $\Lambda_0=\{\nu_1, \ldots, \nu_6\}$,  1-cells {in} $\Lambda_1=\{\eta_1, \ldots, \eta_8\}$, and 2-cells {in} $\Lambda_2=\{\gamma_1, \gamma_2, \gamma_3\}$: (a) non-oriented complex, with cell coefficients in $\Z_2=\{0,1\}$; (b) oriented complex, with cell coefficients in $G=\{-1,0,+1\}$; (c) oriented complex, with cell coefficients in $\R$, using different colors for the maps from $\Lambda_0$,  $\Lambda_1$, and  $\Lambda_2$ to $\R$. To interpret the real numbers here, see Example~\ref{examplee}.}
\label{fig:ex0-abc}
\end{figure}

{
\begin{example}[Boundary]\label{exampled}
The boundary operators are maps $C_p\to C_{p-1}$, with $1\leq p\leq d$. Hence for a 2-complex we have two operators, denoted as $\partial_2: C_2 \to C_1$ and $\partial_1: C_1 \to C_0$, respectively. 
Since they are linear maps between linear spaces, may be represented by matrices of coefficients $[\partial_2]$ and $[\partial_1]$ from the underlying field $\mathbb{F}$. For the unsigned and the signed case (Figures~\ref{fig:ex0-abc}a and ~\ref{fig:ex0-abc}b) we have, respectively:
\[
[\partial_2] = {\scriptsize\mat{
1 & 0 & 0\\
1 & 0 & 0\\
1 & 0 & 1\\
0 & 1 & 0\\
0 & 1 & 1\\
0 & 0 & 1\\
0 & 1 & 0\\
0 & 0 & 1\\
}}\ ,
\ \mbox{{and}}\quad
[\partial'_2] = {\scriptsize\mat{
1 & 0 & 0\\
-1 & 0 & 0\\
1 & 0 & -1\\
0 & 1 & 0\\
0 & -1 & 1\\
0 & 0 & -1\\
0 & 1 & 0\\
0 & 0 & 1\\
}} .
\]

Analogously, for the unsigned $\partial_1$ and the signed $\partial_1'$ operators we have:

\[
\arraycolsep=2.9pt
[\partial_1] = \footnotesize\mat{
1 & 1 & 0 & 0 & 0 & 0 & 0 & 0\\
1 & 0 & 1 & 1 & 1 & 0 & 0 & 0\\
0 & 1 & 1 & 0 & 0 & 1 & 0 & 0\\
0 & 0 & 0 & 1 & 0 & 0 & 1 & 0\\
0 & 0 & 0 & 0 & 1 & 0 & 1 & 1\\
0 & 0 & 0 & 0 & 0 & 1 & 0 & 1\\
}\ ,
\]
and
\[
\arraycolsep=2.9pt
[\partial'_1] = \footnotesize\mat{
-1 & -1 & 0 & 0 & 0 & 0 & 0 & 0\\
1 & 0 & -1 & -1 & -1 & 0 & 0 & 0\\
0 & 1 & 1 & 0 & 0 & -1 & 0 & 0\\
0 & 0 & 0 & 1 & 0 & 0 & -1 & 0\\
0 & 0 & 0 & 0 & 1 & 0 & 1 & -1\\
0 & 0 & 0 & 0 & 0 & 1 & 0 & 1\\
}\ .
\]
As a check, let us compute (a) the 0-boundary of the coordinate representations of the unsigned 1-chains $[d] = [0,1,1,0,1,0,0,0]^t$ (see Figure~\ref{fig:ex0-abc}a) and (b) the signed 1-chain $[d'] = [0,1,-1,0,1,0,0,0]^t$ (see Figure~\ref{fig:ex0-abc}b)
\[
\partial_1 d = [\partial_1][d]\ {mod\ 2} = [1,0,0,0,1,0]^t = \nu_1 + \nu_5 \in C_0,
\]
where the matrix product is computed $mod\ 2$, and
\[
\partial'_1 d' = [\partial'_1][d'] = [-1,0,0,0,1,0]^t = \nu_5 - \nu_1 \in C'_0.
\]
\end{example}
}

\begin{example}[Cell with a hole] \label{ex:example0}

Figure~\ref{fig:squares} shows an example of 2D cellular complex $X=X_2$, comprised of $8$ unit 0-chains (0-cells) $u_0^h$,  $8$ unit 1-chains (1-cells)  $u_1^k$, and 2 unit 2-chains (2-cells)  $u_2^j$. 
\begin{figure}[htbp] 
   \centering
   \includegraphics[width=0.4\linewidth]{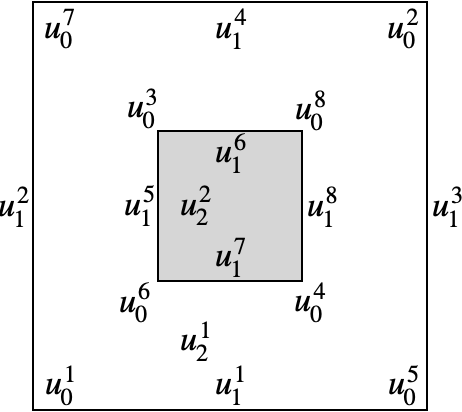} 
   \caption{Cellular 2-complex with two 2-cells, eight 1-cells, and eight 0-cells.}
   \label{fig:squares}
\end{figure}

A user-readable representation of the geometric complex $(X_2,\nu)$ is given below. $\texttt{V}$ is the array of vertices, that provides the embedding map $C_0\to\E^2$, implemented as array $\nu: \N\to\R^2$. \texttt{EV} and \texttt{FV} respectively provide the \emph{canonical} (sorted) LAR  of 1-cells and 2-cells as lists of lists of 0-cells indices. These can be interpreted as user-readable CSR (Compressed Sparse Row) characteristic matrices $M_1$ and $M_2$ of the 0-generators of 1-cells and 2-cells, respectively, according to~\cite{Dicarlo:2014:TNL:2543138.2543294}.

{\scriptsize
\begin{verbatim}
V = [[0.,0.],[3.,3.],[1.,2.],[2.,1.],[3.,0.],[1.,1.],[0.,3.],[2.,2.]]
FV = [[1,2,3,4,5,6,7,8],[3,4,6,8]]
EV = [[1,5],[1,7],[2,5],[2,7],[3,6],[3,8],[4,6],[4,8]]
\end{verbatim}}

The unsigned matrix of the boundary operator $\partial_2 : C_2\to C_1$, computed by filtering elements of value 2 in the matrix $M_1 M_2^t$, is
\begin{eqnarray*}
&&[\partial_2] = \mbox{filter}\,(M_1 M_2^t,2) \\
&&= 
\mbox{filter}\,{\scriptsize\left[\ 
\mat{
1 & 0 & 0 & 0 & 1 & 0 & 0 & 0\\
1 & 0 & 0 & 0 & 0 & 0 & 1 & 0\\
0 & 1 & 0 & 0 & 1 & 0 & 0 & 0\\
0 & 1 & 0 & 0 & 0 & 0 & 1 & 0\\
0 & 0 & 1 & 0 & 0 & 1 & 0 & 0\\
0 & 0 & 1 & 0 & 0 & 0 & 0 & 1\\
0 & 0 & 0 & 1 & 0 & 1 & 0 & 0\\
0 & 0 & 0 & 1 & 0 & 0 & 0 & 1}\,\ 
\mat{
1 & 0\\
1 & 0\\
1 & 1\\
1 & 1\\
1 & 0\\
1 & 1\\
1 & 0\\
1 & 1\\
}\,,\ 2\right]}\\
&&= {\scriptsize
\mat{
1 & 1 & 1 & 1 & 1 & 1 & 1 & 1\\
 0 & 0 & 0 & 0 & 1 & 1 & 1 & 1\\
}^t} \nonumber
\end{eqnarray*}
where the first column represents the non-convex 2-cell with the hole, and the second column  represents the convex cell within the hole. The reader may easily check that the four ones in positions from fifth to eighth in the second column of $[\partial_2]$ correspond to the last four unit 1-chains in \texttt{EV} array.
By multiplication $(\mbox{mod 2})$ of $[\partial_2]$ times the coordinate representation $[c]$ of the 2-complex in Figure~\ref{fig:squares}, \emph{i.e.}, times the \emph{total} 2-chain $ c = u_2^1 + u_2^2 = \vet{1 & 1}^t$, we get the coordinate representation
\[
[\partial_2][c] = 
\arraycolsep=2.9pt
\vet{1 & 1& 1& 1& 0& 0& 0& 0}^t
\]
of the 1-boundary of  $c$, \emph{i.e.}, the cycle $u_1^1 + u_1^2 + u_1^3 + u_1^4$ made by the first four 1-cells in \texttt{EV}. 
\end{example}

\begin{example}[Coboundary]\label{examplee}
The coboundary operator $\delta^p : C^p \to C^{p+1}$  acts on $p$-cochains as the dual of the boundary operator $\partial_{p+1}$ on ($p+1$)-chains. For all $\phi^p \in C^p$ and $c_{p+1} \in C_{p+1}$:
\[
\langle \delta^p \phi^p, c_{p+1} \rangle = \langle \phi^p, \partial_{p+1} c_{p+1} \rangle.
\]

 Recalling that chain-cochain duality means integration, the reader will recognize this defining property as the combinatorial archetype of Stokes' theorem. 
See also that in Figure~\ref{fig:ex0-abc} we have $(\delta^1 \circ \delta^0)(\gamma_1)=0$. This property, \emph{i.e.}, $\delta \circ \delta = 0$,  is the discrete archetype of the fact that the curl of gradient is zero. 
Note that a scalar field, in the discrete version, becomes a real valued 0-cochain to be valued on 0-chains, \emph{i.e.}, on 0-cells.
 It is possible to see~\cite{DiCarlo:2009:DPU:1629255.1629273} that, since we use dual bases, matrices representing dual operators are the transpose of each other: for all $p = 0,\ldots,d-1$, 
\[
  [\delta^p]^t = [\partial_{p+1}] .
\]

In Figure~\ref{fig:ex0-abc}c, coefficients from $\R$ are associated to elementary cochains, as resulting from the evaluation of cochain functions on lower order basis chains. 
When cochain coefficients are taken from $G=\{-1,0,+1\}$ we have coboundary matrices 
$
[\delta^1] = [\partial'_2]^t$  and $[\delta^0] = [\partial'_1]^t
$,
so that, with $\phi = [0,0,0,0,1,0,0,0]\in C^1\equiv C_1$, we get
$
[\delta^1][\phi]^t = [0,-1,1]^t = \gamma_3-\gamma_2 \in C^2\equiv C_2,
$ 
as you can check on Figure~\ref{fig:ex0-abc}b. 
\end{example}

\begin{example}[Application: geographical maps]
\label{example2}

Let us consider a geographical map as the plane arrangement $\mathcal{A}(\mathcal{S})$ generated by a quasi-disjoint set of regions ($2$-complex $\mathcal{R}^2$) superimposed with a road network ($1$-complex $\mathcal{R}^1$).
To this purpose, we take as input the collection of data $\mathcal{S} = \{\mathcal{R}^2,\mathcal{R}^1\}$, and select the combinatorial union of their $1$-skeletons $\mathcal{S}_1 = \mathcal{R}^2_1 \cup \mathcal{S}^1_1$. From this set of $1$-cells --- which is not a $1$-complex, since cells may intersect away from their boundary vertices --- we compute the cellular complex $X_2 = \mathcal{A}(\mathcal{S}_1)$ and the associated chain complex $C_\bullet$.
Here, 1D roads are simply a particular chain in the linear space $C_1$, whereas each region is a chain in $C_2$, \emph{i.e.}, a sum of basis $2$-chains. The length of any portion of road is the real number attached by a $1$-cochain to that particular 1-chain. Analogously, the area of a region is the real number produced by a $2$-cochain evaluated on that $2$-chain. By linearity, the number is the sum of areas of basis 2-chains it is linear combination of.
 
\end{example}

\section*{References}

\bibliography{gmp2019}

\begin{thebibliography}{100}
\expandafter\ifx\csname url\endcsname\relax
  \def\url#1{\texttt{#1}}\fi
\expandafter\ifx\csname urlprefix\endcsname\relax\def\urlprefix{URL }\fi
\expandafter\ifx\csname href\endcsname\relax
  \def\href#1#2{#2} \def\path#1{#1}\fi

\bibitem{Edelsbrunner:95}
C.~Delfinado, H.~Edelsbrunner, An incremental algorithm for {Betti} numbers of
  sinplicial complexes on the 3-sphere, Computer Aided Geometric Design 12
  (1995) 771--784.

\bibitem{hatcher:2002}
A.~Hatcher, Algebraic topology, Cambridge University Press, 2002.

\bibitem{Rourke:Sanderson:1982}
C.~Rourke, B.~Sanderson,
  \href{http://doi.org/10.1007/978-3-642-81735-9}{Introduction to
  Piecewise-Linear Topology}, Springer-Verlag, Berlin, Heidelberg, 1982.
\newline\urlprefix\url{http://doi.org/10.1007/978-3-642-81735-9}

\bibitem{Alayrangues2015}
S.~Alayrangues, G.~Damiand, P.~Lienhardt, S.~Peltier, Homology of cellular
  structures allowing multi-incidence, Discrete {\&} Computational Geometry
  54~(1) (2015) 42--77.
\newblock \href {http://dx.doi.org/10.1007/s00454-015-9662-5}
  {\path{doi:10.1007/s00454-015-9662-5}}.

\bibitem{Lienhardt:2014}
G.~Damiand, P.~Lienhardt, Combinatorial Maps: Efficient Data Structures for
  Computer Graphics and Image Processing, {CRC} Press, 2014.

\bibitem{Munkres:84}
J.~Munkres, Elements of Algebraic Topology, Addison-Wesley, Reading, {MA},
  1984.

\bibitem{Kannan:79}
R.~Kannan, A.~Bachem, \href{https://doi.org/10.1137/0208040}{Polynomial
  algorithms for computing the smith and hermite normal forms of an integer
  matrix}, {SIAM} J. Comput.~(8) (1979) 1979.
\newline\urlprefix\url{https://doi.org/10.1137/0208040}

\bibitem{Paoluzzi:1993:DMS:169728.169719}
A.~Paoluzzi, F.~Bernardini, C.~Cattani, V.~Ferrucci, Dimension-independent
  modeling with simplicial complexes, {ACM} Trans. Graph. 12~(1) (1993)
  56--102.
\newblock \href {http://dx.doi.org/10.1145/169728.169719}
  {\path{doi:10.1145/169728.169719}}.

\bibitem{fhktww-a-07}
E.~Fogel, D.~Halperin, L.~Kettner, M.~Teillaud, R.~Wein, N.~Wolpert,
  Arrangements, in: J.-D. Boissonat, M.~Teillaud (Eds.), Effective
  Computational Geometry for Curves and Surfaces, Mathematics and
  Visualization, Springer, 2007, Ch.~1, pp. 1--66.

\bibitem{Fabri:2000:DCC:358668.358687}
A.~Fabri, G.-J. Giezeman, L.~Kettner, S.~Schirra, S.~Sch\"{o}nherr, On the
  design of cgal a computational geometry algorithms library, Softw. Pract.
  Exper. 30~(11) (2000) 1167--1202.
\newblock \href
  {http://dx.doi.org/10.1002/1097-024X(200009)30:11<1167::AID-SPE337>3.0.CO;2-B}
  {\path{doi:10.1002/1097-024X(200009)30:11<1167::AID-SPE337>3.0.CO;2-B}}.

\bibitem{Hachenberger:2007:BOS:1247750.1248141}
P.~Hachenberger, L.~Kettner, K.~Mehlhorn,
  \href{http://dx.doi.org/10.1016/j.comgeo.2006.11.009}{Boolean operations on
  3d selective nef complexes: Data structure, algorithms, optimized
  implementation and experiments}, Comput. Geom. Theory Appl. 38~(1-2) (2007)
  64--99.
\newblock \href {http://dx.doi.org/10.1016/j.comgeo.2006.11.009}
  {\path{doi:10.1016/j.comgeo.2006.11.009}}.
\newline\urlprefix\url{http://dx.doi.org/10.1016/j.comgeo.2006.11.009}

\bibitem{bieri:95}
H.~Bieri, Nef polyhedra: A brief introduction, in: H.~Hagen, G.~Farin,
  H.~Noltemeier (Eds.), Geometric Modelling, Springer Vienna, Vienna, 1995, pp.
  43--60.

\bibitem{Goodman:2017:HDC:285869}
J.~E. Goodman, J.~O'Rourke, C.~D. T\`oth (Eds.), Handbook of Discrete and
  Computational Geometry -- Third Edition, CRC Press, Inc., Boca Raton, FL,
  USA, 2017.

\bibitem{Ziegler:92}
A.~Bj\"orner, G.~Ziegler, Combinatorial stratification of complex arrangements,
  J. Amer. Math. Soc.~(5) (1992) 105--149.

\bibitem{Edelsbrunner:1987:ACG:28905}
H.~Edelsbrunner, Algorithms in Combinatorial Geometry, Springer-Verlag New
  York, Inc., New York, NY, {USA}, 1987.

\bibitem{Birkhoff:1948}
G.~Birkhoff, Lattice Theory (Revised ed.), AMS Colloquium publications,
  American Mathematical Society, New York, {NY}, 1948.

\bibitem{ieee-tase}
A.~DiCarlo, F.~Milicchio, A.~Paoluzzi, V.~Shapiro, Chain-based representations
  for solid and physical modeling, Automation Science and Engineering, {IEEE}
  Transactions on 6~(3) (2009) 454 --467.
\newblock \href {http://dx.doi.org/10.1109/TASE.2009.2021342}
  {\path{doi:10.1109/TASE.2009.2021342}}.

\bibitem{Baumgart:1972:WEP:891970}
B.~G. Baumgart, Winged edge polyhedron representation., Tech. Rep. Stan-CS-320,
  Stanford, CA, {USA} (1972).

\bibitem{Muller:78}
D.~E. Muller, F.~P. Preparata, Finding the intersection of two convex
  polyhedra, Theoretical Computer Science 7 (1978) 217–236.

\bibitem{Dobkin:1987:PMT:41958.41967}
D.~P. Dobkin, M.~J. Laszlo, Primitives for the manipulation of
  three-dimensional subdivisions, in: Proceedings of the third annual symposium
  on Computational geometry, SCG '87, {ACM}, New York, NY, {USA}, 1987, pp.
  86--99.
\newblock \href {http://dx.doi.org/10.1145/41958.41967}
  {\path{doi:10.1145/41958.41967}}.

\bibitem{Hoffmann:1987:RSO:866286}
C.~M. Hoffmann, J.~E. Hopcroft, M.~S. Karasick, Robust set operations on
  polyhedral solids, Tech. rep., Ithaca, NY, USA (1987).

\bibitem{Rossignac:89}
J.~R. Rossignac, M.~A. O'Connor, A dimension-independent model for pointsets
  with internal structures and incomplete boundaries, Tech. rep., IBM Research
  Division, Yorktown Heights, N.Y. 10598 (89).

\bibitem{Woodwark:99}
C.~Armstrong, A.~Bowyer, S.~Cameron, J.~Corney, G.~Jared, R.~Martin,
  A.~Middleditch, M.~Sabin, J.~Salmon, J.~Woodwark, Djinn: a geometric
  interface for solid modelling, Tech. rep., Information Geometers Ltd.,
  Winchester, {UK} (1999).

\bibitem{Middleditch:92}
A.~Middleditch, Cellular models of mixed dimension, Tech. Rep. BRU/CAE/92:3,
  Brunel University Centre for Geometric Modelling and Design (April 1992).

\bibitem{Zhou:2016:MAS:2897824.2925901}
Q.~Zhou, E.~Grinspun, D.~Zorin, A.~Jacobson, Mesh arrangements for solid
  geometry, ACM Trans. Graph. 35~(4) (2016) 39:1--39:15.
\newblock \href {http://dx.doi.org/10.1145/2897824.2925901}
  {\path{doi:10.1145/2897824.2925901}}.

\bibitem{4055948}
K.~Weiler, Edge-based data structures for solid modeling in curved-surface
  environments, Computer Graphics and Applications, {IEEE} 5~(1) (1985) 21
  --40.
\newblock \href {http://dx.doi.org/10.1109/MCG.1985.276271}
  {\path{doi:10.1109/MCG.1985.276271}}.

\bibitem{Ala:1992:PAB:616022.617736}
S.~R. Ala, Performance anomalies in boundary data structures, {IEEE} Comput.
  Graph. Appl. 12~(2) (1992) 49--58.
\newblock \href {http://dx.doi.org/10.1109/38.124288}
  {\path{doi:10.1109/38.124288}}.

\bibitem{bowyer1995introducing}
A.~Bowyer, G.~M. Society,
  \href{http://books.google.it/books?id=oCGGAAAACAAJ}{Introducing {Djinn}: A
  Geometric Interface for Solid Modelling}, Information Geometers [for] the
  Geometric Modelling Society, 1995.
\newline\urlprefix\url{http://books.google.it/books?id=oCGGAAAACAAJ}

\bibitem{bowyer1995svlis}
A.~Bowyer, \href{http://books.google.it/books?id=hYqwAAAACAAJ}{SvLis
  Set-theoretic Kernel Modeller: Introduction and User Manual}, Information
  Geometers, 1995.
\newline\urlprefix\url{http://books.google.it/books?id=hYqwAAAACAAJ}

\bibitem{Braid:1975:SSB:360715.360727}
I.~C. Braid, The synthesis of solids bounded by many faces, Commun. {ACM}
  18~(4) (1975) 209--216.
\newblock \href {http://dx.doi.org/10.1145/360715.360727}
  {\path{doi:10.1145/360715.360727}}.

\bibitem{Brisson:1989:RGS:73833.73858}
E.~Brisson, Representing geometric structures in $d$ dimensions: topology and
  order, in: Proc.~of the 5-th Annual Symposium on Computational Geometry, SCG
  '89, Acm, New York, NY, {USA}, 1989, pp. 218--227.
\newblock \href {http://dx.doi.org/10.1145/73833.73858}
  {\path{doi:10.1145/73833.73858}}.

\bibitem{cadanda}
C.~Bajaj, A.~Paoluzzi, G.~Scorzelli, Progressive conversion from b-rep to bsp
  for streaming geometric modeling, Computer-Aided Design and Applications
  3~(5-6).

\bibitem{Gomes:1999:MMB:304012.304039}
A.~Gomes, A.~Middleditch, C.~Reade, A mathematical model for boundary
  representations of n-dimensional geometric objects, in: Procs of the fifth
  ACM Symp.~on Solid modeling and applications, SMA '99, {ACM}, New York, NY,
  {USA}, 1999, pp. 270--277.
\newblock \href {http://dx.doi.org/10.1145/304012.304039}
  {\path{doi:10.1145/304012.304039}}.

\bibitem{Guibas:1985:PMG:282918.282923}
L.~Guibas, J.~Stolfi, Primitives for the manipulation of general subdivisions
  and the computation of voronoi, {ACM} Trans. Graph. 4~(2) (1985) 74--123.
\newblock \href {http://dx.doi.org/10.1145/282918.282923}
  {\path{doi:10.1145/282918.282923}}.

\bibitem{HoffmannK01}
C.~M. Hoffmann, K.-J. Kim, Towards valid parametric cad models, Computer-Aided
  Design 33~(1) (2001) 81--90.
\newblock \href
  {http://dx.doi.org/http://dx.doi.org/10.1016/S0010-4485(00)00073-7}
  {\path{doi:http://dx.doi.org/10.1016/S0010-4485(00)00073-7}}.

\bibitem{Kalay:1989:HET:63718.63719}
Y.~E. Kalay, The hybrid edge: a topological data structure for vertically
  integrated geometric modelling, Comput. Aided Des. 21~(3) (1989) 130--140.
\newblock \href {http://dx.doi.org/10.1016/0010-4485(89)90067-5}
  {\path{doi:10.1016/0010-4485(89)90067-5}}.

\bibitem{Lee:2001:PES:376957.376976}
S.~H. Lee, K.~Lee, Partial entity structure: a compact non-manifold boundary
  representation based on partial topological entities, in: Proceedings of the
  sixth ACM symposium on Solid modeling and applications, SMA '01, Acm, New
  York, NY, {USA}, 2001, pp. 159--170.
\newblock \href {http://dx.doi.org/10.1145/376957.376976}
  {\path{doi:10.1145/376957.376976}}.

\bibitem{Lienhardt:1991:TMB:115604.115610}
P.~Lienhardt, Topological models for boundary representation: a comparison with
  n-dimensional generalized maps, Comput. Aided Des. 23~(1) (1991) 59--82.
\newblock \href {http://dx.doi.org/10.1016/0010-4485(91)90082-8}
  {\path{doi:10.1016/0010-4485(91)90082-8}}.

\bibitem{Mantyla:1988:ISM:60949}
M.~Mantyla, Introduction to Solid Modeling, W. H. Freeman \& Co., New York, NY,
  {USA}, 1988.

\bibitem{Paoluzzi:1989:BAO:70248.70249}
A.~Paoluzzi, M.~Ramella, A.~Santarelli, Boolean algebra over linear polyhedra,
  Comput. Aided Des. 21~(10) (1989) 474--484.
\newblock \href {http://dx.doi.org/10.1016/0010-4485(89)90055-9}
  {\path{doi:10.1016/0010-4485(89)90055-9}}.

\bibitem{Paoluzzi:1995:GPP:212332.212349}
A.~Paoluzzi, V.~Pascucci, M.~Vicentino, Geometric programming: a programming
  approach to geometric design, {ACM} Trans. Graph. 14~(3) (1995) 266--306.
\newblock \href {http://dx.doi.org/10.1145/212332.212349}
  {\path{doi:10.1145/212332.212349}}.

\bibitem{Pascucci:1995:DCB:218013.218055}
V.~Pascucci, V.~Ferrucci, A.~Paoluzzi, Dimension-independent convex-cell based
  {HPC}: representation scheme and implementation issues, in: Proceedings of
  the third ACM Symposium on Solid Modeling and Applications, SMA '95, Acm, New
  York, NY, {USA}, 1995, pp. 163--174.
\newblock \href {http://dx.doi.org/10.1145/218013.218055}
  {\path{doi:10.1145/218013.218055}}.

\bibitem{Pratt94ashape}
M.~J. Pratt, B.~D. Anderson, A shape modelling api for the {STEP} standard, in:
  in Fourteenth Int.~Conference on Atomic Physics, 1994, pp. 1--7.

\bibitem{Raghothama:1999:CUD:304012.304019}
S.~Raghothama, V.~Shapiro, Consistent updates in dual representation systems,
  in: Proceedings of the fifth ACM symposium on Solid modeling and
  applications, SMA '99, Acm, New York, NY, {USA}, 1999, pp. 65--75.
\newblock \href {http://dx.doi.org/10.1145/304012.304019}
  {\path{doi:10.1145/304012.304019}}.

\bibitem{Rap97}
A.~Rappoport, The generic geometric complex (ggc): a modeling scheme for
  families of decomposed pointsets, in: Proceedings of the fourth ACM symposium
  on Solid modeling and applications, SMA '97, Acm, New York, NY, {USA}, 1997,
  pp. 19--30.
\newblock \href {http://dx.doi.org/10.1145/267734.267749}
  {\path{doi:10.1145/267734.267749}}.

\bibitem{Requicha:1980:RRS:356827.356833}
A.~G. Requicha, Representations for rigid solids: Theory, methods, and systems,
  {ACM} Comput. Surv. 12~(4) (1980) 437--464.
\newblock \href {http://dx.doi.org/10.1145/356827.356833}
  {\path{doi:10.1145/356827.356833}}.

\bibitem{RequichaVoelcker:77}
A.~A.~G. Requicha, H.~B. Voelcker, Constructive solid geometry, Tech. Rep.
  TM-25, Production Automation Project, Univ. of Rochester (1977).

\bibitem{Rossignac:1991:CNG:115604.115606}
J.~R. Rossignac, A.~A.~G. Requicha,
  \href{http://dx.doi.org/10.1016/0010-4485(91)90078-B}{Constructive
  non-regularized geometry}, Comput. Aided Des. 23~(1) (1991) 21--32.
\newblock \href {http://dx.doi.org/10.1016/0010-4485(91)90078-B}
  {\path{doi:10.1016/0010-4485(91)90078-B}}.
\newline\urlprefix\url{http://dx.doi.org/10.1016/0010-4485(91)90078-B}

\bibitem{Rossignac:SGC:90}
J.~R. Rossignac, M.~A. O'Connor, {SGC:} a dimension-independent model for
  pointsets with internal structures and incomplete boundaries, in: Geometric
  modeling for product engineering, North-Holland, 1990.

\bibitem{Shapiro:1991:RSS:124951}
V.~Shapiro, Representations of semi-algebraic sets in finite algebras generated
  by space decompositions, Ph.D. thesis, Ithaca, NY, {USA}, uMI Order No.
  GAX91-31407 (1991).

\bibitem{Shapiro:1995:PFS:218013.218029}
V.~Shapiro, D.~L. Vossler, What is a parametric family of solids?, in: Proc.~of
  the third ACM Symp.~on Solid modeling and applications, SMA '95, ACM, 1995,
  pp. 43--54.
\newblock \href {http://dx.doi.org/10.1145/218013.218029}
  {\path{doi:10.1145/218013.218029}}.

\bibitem{Silva:81}
C.~E. Silva, Alternative definitions of faces in boundary representations of
  solid objects, Tech. Rep. TM-36, Production Automation Project, Univ. of
  Rochester (1981).

\bibitem{Weiler:86}
K.~J. Weiler, Topological structures for geometric modeling, Ph.D. thesis,
  Rensselaer Polytechnic Institute (1986).

\bibitem{Weiler:88}
K.~J. Weiler, The radial edge structure: A topological representation for
  non-manifold geometric modelling, in: M.~Wozny, H.~McLaughlin, J.~Encarnacao
  (Eds.), Geometric modelling for CAD applications, Amsterdam, 1988, pp. 3--12.

\bibitem{Woo:85}
T.~Woo, A combinatorial analysis of boundary data structure schemata, Computer
  Graphics \& Applications, {IEEE} 5~(3) (1985) 19--27.

\bibitem{wozny1990geometric}
M.~Wozny, J.~Turner, K.~Preiss,
  \href{http://books.google.it/books?id=-BkfAQAAIAAJ}{Geometric modeling for
  product engineering: IFIP WG 5.2/NSF Working Conf. on Geometric Modeling,
  Rensselaerville, U.S.A., 18-22 September, 1988}, North-Holland, 1990.
\newline\urlprefix\url{http://books.google.it/books?id=-BkfAQAAIAAJ}

\bibitem{Yamaguchi:85}
F.~Yamaguchi, T.~Tokieda, Bridge edge and triangulation approach in solid
  modeling, in: T.~Kunii (Ed.), Frontiers in Computer Graphics, Springer
  Verlag, Berlin, 1985.

\bibitem{yamaguchi1995ntb}
Y.~Yamaguchi, F.~Kimura, Nonmanifold topology based on coupling entities,
  Computer Graphics and Applications, {IEEE} 15~(1) (1995) 42--50.

\bibitem{Hoffmann:91}
C.~M. Hoffmann, G.~Vanĕček, Fundamental techniques for geometric and solid
  modeling, Tech. Rep. 91-044, Purdue University (1991).

\bibitem{Hoffmann:1989:GSM:74803}
C.~M. Hoffmann, Geometric and Solid Modeling: An Introduction, Morgan Kaufmann
  Publishers Inc., San Francisco, CA, USA, 1989.

\bibitem{PALMER1995733}
R.~S. Palmer, Chain models and finite element analysis: An executable chains
  formulation of plane stress, Computer Aided Geometric Design 12~(7) (1995)
  733 -- 770, grid Generation, Finite Elements, and Geometric Design.
\newblock \href
  {http://dx.doi.org/https://doi.org/10.1016/0167-8396(95)00015-X}
  {\path{doi:https://doi.org/10.1016/0167-8396(95)00015-X}}.

\bibitem{Palmer1993}
R.~S. Palmer, V.~Shapiro, Chain models of physical behavior for engineering
  analysis and design, Research in Engineering Design 5~(3) (1993) 161--184.
\newblock \href {http://dx.doi.org/10.1007/BF01608361}
  {\path{doi:10.1007/BF01608361}}.

\bibitem{Hirani:2003:DEC:959640}
A.~N. Hirani, Discrete exterior calculus, Ph.D. thesis, Pasadena, CA, USA,
  aAI3086864 (2003).

\bibitem{Desbrun:2006:DDF:1185657.1185665}
M.~Desbrun, E.~Kanso, Y.~Tong, Discrete differential forms for computational
  modeling, in: ACM SIGGRAPH 2006 Courses, SIGGRAPH '06, ACM, New York, NY,
  USA, 2006, pp. 39--54.
\newblock \href {http://dx.doi.org/10.1145/1185657.1185665}
  {\path{doi:10.1145/1185657.1185665}}.

\bibitem{Elcott:2006:BYO:1185657.1185666}
S.~Elcott, P.~Schroder, Building your own dec at home, in: ACM SIGGRAPH 2006
  Courses, SIGGRAPH '06, ACM, New York, NY, USA, 2006, pp. 55--59.
\newblock \href {http://dx.doi.org/10.1145/1185657.1185666}
  {\path{doi:10.1145/1185657.1185666}}.

\bibitem{arnold_falk_winther_2006}
D.~N. Arnold, R.~S. Falk, R.~Winther, Finite element exterior calculus,
  homological techniques, and applications, Acta Numerica 15 (2006) 1–155.
\newblock \href {http://dx.doi.org/10.1017/S0962492906210018}
  {\path{doi:10.1017/S0962492906210018}}.

\bibitem{Arnold:2010}
D.~N. Arnold, R.~S. Falk, R.~Winther, Finite element exterior calculus: from
  {H}odge theory to numerical stability, Bull. Amer. Math. Soc. (N.S.) 47
  (2010) 281--354.

\bibitem{Arnold:2018}
D.~N. Arnold, Finite Element Exterior Calculus, Vol.~93 of CBMS-NSF Regional
  Conference Series in Applied Mathematics, Society for Industrial and Applied
  Mathematics (SIAM), Philadelphia, PA, 2018.

\bibitem{Tonti:1975}
E.~Tonti, On the formal structure of physical theories, Tech. rep., Italian
  National Research Council (1975).

\bibitem{Tonti:2013}
E.~Tonti, The Mathematical Structure of Classical and Relativistic Physics,
  Birkh\"auser, 2013.

\bibitem{Ferretti:2014}
E.~Ferretti, The Cell Method: A Purely Algebraic Computational Method in
  Physics and Engineering, Momentum Press, 2014.

\bibitem{DiCarlo:2009:DPU:1629255.1629273}
A.~DiCarlo, F.~Milicchio, A.~Paoluzzi, V.~Shapiro, Discrete physics using
  metrized chains, in: 2009 SIAM/ACM Joint Conference on Geometric and Physical
  Modeling, SPM '09, Acm, New York, NY, {USA}, 2009, pp. 135--145.
\newblock \href
  {http://arxiv.org/abs/http://doi.acm.org/10.1145/1629255.1629273}
  {\path{arXiv:http://doi.acm.org/10.1145/1629255.1629273}}, \href
  {http://dx.doi.org/10.1145/1629255.1629273}
  {\path{doi:10.1145/1629255.1629273}}.

\bibitem{Dicarlo:2014:TNL:2543138.2543294}
A.~Dicarlo, A.~Paoluzzi, V.~Shapiro, Linear algebraic representation for
  topological structures, Comput. Aided Des. 46 (2014) 269--274.
\newblock \href {http://dx.doi.org/10.1016/j.cad.2013.08.044}
  {\path{doi:10.1016/j.cad.2013.08.044}}.

\bibitem{Bell:SpMV:NVIDIA:2008}
N.~Bell, M.~Garland, Efficient sparse matrix-vector multiplication on {CUDA},
  NVIDIA Technical Report NVR-2008-004, NVIDIA Corporation (Dec. 2008).

\bibitem{gemmexp}
A.~Bulu\c{c}, J.~R. Gilbert, Parallel sparse matrix-matrix multiplication and
  indexing: Implementation and experiments, {SIAM} Journal of Scientific
  Computing (SISC) 34~(4) (2012) 170 -- 191.
\newblock \href {http://dx.doi.org/10.1137/110848244}
  {\path{doi:10.1137/110848244}}.

\bibitem{Davis:2006:DMS:1196434}
T.~A. Davis, Direct Methods for Sparse Linear Systems (Fundamentals of
  Algorithms 2), Society for Industrial and Applied Mathematics, Philadelphia,
  PA, {USA}, 2006.

\bibitem{BEKS14}
J.~Bezanson, A.~Edelman, S.~Karpinski, V.~B. Shah, Julia: A fresh approach to
  numerical computing, {SIAM} Review 59~(1) (2017) 65--98.
\newblock \href {http://arxiv.org/abs/http://dx.doi.org/10.1137/141000671}
  {\path{arXiv:http://dx.doi.org/10.1137/141000671}}, \href
  {http://dx.doi.org/10.1137/141000671} {\path{doi:10.1137/141000671}}.

\bibitem{Coxeter:1967}
H.~S.~M. Coxeter, S.~L. Greitzer, Geometry Revisited, Math. Assoc. Amer.,
  Washington, D.C., 1967.

\bibitem{Baladze:EoM}
D.~Baladze, \href{http://www.encyclopediaofmath.org/}{Cw-complex}, in:
  Encyclopedia of Mathematics, Springer \& European Mathematical Society, 2012.
\newline\urlprefix\url{http://www.encyclopediaofmath.org/}

\bibitem{Cormen:2009:IAT:1614191}
T.~H. Cormen, C.~E. Leiserson, R.~L. Rivest, C.~Stein, Introduction to
  Algorithms, Third Edition, 3rd Edition, The {MIT} Press, 2009.

\bibitem{Jarvis:1973:ICH}
R.~A. Jarvis, On the identification of the convex hull of a finite set of
  points in the plane, Information Processing Letters 2~(1) (1973) 18--21.

\bibitem{vialar2016handbook}
T.~Vialar, \href{https://books.google.com/books?id=-tcVMQAACAAJ}{Handbook of
  Mathematics}, HDBoM, 2016, (Def 805).
\newline\urlprefix\url{https://books.google.com/books?id=-tcVMQAACAAJ}

\bibitem{Hopcroft:1973:AEA:362248.362272}
J.~Hopcroft, R.~Tarjan, Algorithm 447: Efficient algorithms for graph
  manipulation, Commun. ACM 16~(6) (1973) 372--378.
\newblock \href {http://dx.doi.org/10.1145/362248.362272}
  {\path{doi:10.1145/362248.362272}}.

\bibitem{Ballard:2015}
G.~Ballard, A.~Druinsky, Sparse matrix-matrix multiplication: Applications,
  algorithms, and implementations, in: SIAM Conference on Applied Linear
  Algebra, Atlanta, {GA}, 2015.

\bibitem{DEHLINGER2014869}
C.~Dehlinger, J.~Dufourd, Formal specification and proofs for the topology and
  classification of combinatorial surfaces, Computational Geometry 47~(9)
  (2014) 869 -- 890.
\newblock \href
  {http://dx.doi.org/https://doi.org/10.1016/j.comgeo.2014.04.007}
  {\path{doi:https://doi.org/10.1016/j.comgeo.2014.04.007}}.

\bibitem{Attene:2014:DRS:2953208.2953514}
M.~Attene, Direct repair of self-intersecting meshes, Graph. Models 76~(6)
  (2014) 658--668.
\newblock \href {http://dx.doi.org/10.1016/j.gmod.2014.09.002}
  {\path{doi:10.1016/j.gmod.2014.09.002}}.

\bibitem{shewchuk96b}
J.~R. Shewchuk, Triangle: {E}ngineering a {2D} {Q}uality {M}esh {G}enerator and
  {D}elaunay {T}riangulator, in: M.~C. Lin, D.~Manocha (Eds.), Applied
  Computational Geometry: Towards Geometric Engineering, Vol. 1148 of Lecture
  Notes in Computer Science, Springer-Verlag, 1996, pp. 203--222.

\bibitem{SHEWCHUK200221}
J.~R. Shewchuk,
  \href{http://www.sciencedirect.com/science/article/pii/S0925772101000475}{Delaunay
  refinement algorithms for triangular mesh generation}, Computational Geometry
  22~(1) (2002) 21 -- 74.
\newblock \href
  {http://dx.doi.org/https://doi.org/10.1016/S0925-7721(01)00047-5}
  {\path{doi:https://doi.org/10.1016/S0925-7721(01)00047-5}}.
\newline\urlprefix\url{http://www.sciencedirect.com/science/article/pii/S0925772101000475}

\bibitem{Shewchuk:1996:RAF:237218.237337}
J.~R. Shewchuk, \href{http://doi.acm.org/10.1145/237218.237337}{Robust adaptive
  floating-point geometric predicates}, in: Proceedings of the Twelfth Annual
  Symposium on Computational Geometry, SCG '96, Acm, New York, NY, {USA}, 1996,
  pp. 141--150.
\newblock \href {http://dx.doi.org/10.1145/237218.237337}
  {\path{doi:10.1145/237218.237337}}.
\newline\urlprefix\url{http://doi.acm.org/10.1145/237218.237337}

\bibitem{Campen:2010}
M.~Campen, L.~Kobbelt, Exact and robust (self-)intersections for polygonal
  meshes, Computer Graphics Forum~(29) (2010) 397--406.

\bibitem{Paoluzzi2003a}
A.~Paoluzzi, \href{https://doi.org/10.1002/0470013885}{Geometric Programming
  for Computer Aided Design}, John Wiley \& Sons, Chichester, {UK}, 2003.
\newline\urlprefix\url{https://doi.org/10.1002/0470013885}

\bibitem{Paoluzzi:2004:PDB:1217875.1217907}
A.~Paoluzzi, V.~Pascucci, G.~Scorzelli,
  \href{http://dl.acm.org/citation.cfm?id=1217875.1217907}{Progressive
  dimension-independent boolean operations}, in: Proceedings of the Ninth ACM
  Symposium on Solid Modeling and Applications, SM '04, Eurographics
  Association, Aire-La-Ville, Switzerland, Switzerland, 2004, pp. 203--211.
\newline\urlprefix\url{http://dl.acm.org/citation.cfm?id=1217875.1217907}

\bibitem{ScorzelliPP-PSM2008}
G.~Scorzelli, A.~Paoluzzi, V.~Pascucci, Parallel solid modeling using bsp
  dataflow, International Journal of Computational Geometry and Applications
  18~(5) (2008) 441--467.

\bibitem{Guibas:98}
L.~J. Guibas, D.~H. Marimont,
  \href{https://doi.org/10.1142/S0218195998000096}{Rounding arrangements
  dynamically}, Int.~Journal of Comp.~Geometry \& Applications 8~(2) (1998)
  157--178.
\newline\urlprefix\url{https://doi.org/10.1142/S0218195998000096}

\bibitem{BARKI20151235}
H.~Barki, G.~Guennebaud, S.~Foufou, Exact, robust, and efficient regularized
  booleans on general 3d meshes, Computers \& Mathematics With Applications
  70~(6) (2015) 1235 -- 1254.
\newblock \href {http://dx.doi.org/https://doi.org/10.1016/j.camwa.2015.06.016}
  {\path{doi:https://doi.org/10.1016/j.camwa.2015.06.016}}.

\bibitem{stag.20151290}
F.~Spini, M.~Sportillo, M.~Virgadamo, E.~Marino, A.~Bottaro, A.~Paoluzzi,
  Hijson: Cartographic document for web modeling of interactive indoor mapping,
  in: A.~Giachetti, S.~Biasotti, M.~Tarini (Eds.), Smart Tools and Apps for
  Graphics - Eurographics Italian Chapter Conference, The Eurographics
  Association, 2015.
\newblock \href {http://dx.doi.org/10.2312/stag.20151290}
  {\path{doi:10.2312/stag.20151290}}.

\bibitem{visigrapp17:cvdlab}
E.~Marino, F.~Spini, D.~Salvati, C.~Vadal\`a, M.~Vicentino, A.~Paoluzzi,
  A.~Bottaro, Modeling semantics for building deconstruction, in: Proc., 12wt
  Int.~Conf.~on Computer Graphics Theory and Applications, 2017.

\bibitem{SpiniMDCP-WEB3D2016.bib}
F.~Spini, E.~Marino, M.~D'Antimi, E.~Carra, A.~Paoluzzi, Web 3d indoor
  authoring and vr exploration via texture baking service, in: 21st annual
  Web3D 2016 Conference, Anaheim, {CA}, 2016.

\bibitem{paoluzziMS:2014}
A.~Paoluzzi, E.~Marino, F.~Spini, {LAR-ABC}, a representation of architectural
  geometry from concept of spaces, to design of building fabric, to
  construction simulation, in: P.~Block, J.~Knippers, N.~J. Mitra, W.~Wang
  (Eds.), Advances in Architectural Geometry 2014, Springer Int.~Publishing,
  2015, pp. 353--372.
\newblock \href {http://dx.doi.org/10.1007/978-3-319-11418-7_23}
  {\path{doi:10.1007/978-3-319-11418-7_23}}.

\bibitem{DBLP:journals/corr/abs-1710-07819}
F.~Furiani, G.~Martella, A.~Paoluzzi,
  \href{http://arxiv.org/abs/1710.07819}{Geometric computing with chain
  complexes: Design and features of a julia package}, CoRR abs/1710.07819.
\newblock \href {http://arxiv.org/abs/1710.07819} {\path{arXiv:1710.07819}}.
\newline\urlprefix\url{http://arxiv.org/abs/1710.07819}

\bibitem{Besard:2017}
T.~Besard,
  \href{https://devblogs.nvidia.com/gpu-computing-julia-programming-language/}{High-performance
  gpu computing in the julia programming language}, Tech. rep., NVIDIA
  Developer Blog (October, 25 2017).
\newline\urlprefix\url{https://devblogs.nvidia.com/gpu-computing-julia-programming-language/}

\bibitem{besard:2017a}
T.~Besard, C.~Foket, B.~De~Sutter, Effective extensible programming: Unleashing
  {Julia} on {GPUs}, IEEE Transactions on Parallel and Distributed Systems\href
  {http://dx.doi.org/10.1109/TPDS.2018.2872064}
  {\path{doi:10.1109/TPDS.2018.2872064}}.

\bibitem{Goodfellow:2016:DL:3086952}
I.~Goodfellow, Y.~Bengio, A.~Courville, Deep Learning, The MIT Press, 2016.

\bibitem{Boxel:2016:DLT:3019358}
D.~V. Boxel, Deep Learning with TensorFlow, Packt Publishing, 2016.

\bibitem{Paoluzzi-DCFJ2016}
A.~Paoluzzi, A.~DiCarlo, F.~Furiani, M.~Jirik, Cad models from medical
  images using lar, Computer-Aided Design and Applications\href
  {http://dx.doi.org/10.1080/16864360.2016.1168216}
  {\path{doi:10.1080/16864360.2016.1168216}}.

\bibitem{ClementiSSPP-CAD16}
G.~Clementi, D.~Salvati, G.~Scorzelli, A.~Paoluzzi, V.~Pascucci, Progressive
  extraction of neural models from high-resolution {3D} images of brain, in:
  13th Int.~Conf.~on CAD \& Applications, Vancouver, {BC}, Canada, 2016.

\bibitem{10.1109/MC.2014.103}
Y.~L. Moon, K.~Sugamoto, A.~Paoluzzi, A.~D. Carlo, J.~Kwak, D.~S. Shin, D.~O.
  Kim, D.~H. Lee, J.~Kim, Standardizing 3d medical imaging, Computer 47~(4)
  (2014) 76--79.
\newblock \href
  {http://dx.doi.org/http://doi.ieeecomputersociety.org/10.1109/MC.2014.103}
  {\path{doi:http://doi.ieeecomputersociety.org/10.1109/MC.2014.103}}.

\bibitem{hinzl:thesis:2007}
R.~Heinzl, Concepts for scientific computing, Ph.D. thesis, Wien, \"Osterreich
  (2007).

\end{thebibliography}

\end{document}